\documentclass[10pt,journal,a4paper,final,oneside,twocolumn]{IEEEtran}
\usepackage{amsmath,amsfonts}
\usepackage{algorithmic}
\usepackage{algorithm}
\usepackage{array}
\usepackage{textcomp}
\usepackage{stfloats}
\usepackage{url}
\usepackage{verbatim}
\usepackage{graphicx}
\usepackage{multirow}
\usepackage{amssymb}
\usepackage{bbding}
\usepackage{setspace}
\usepackage{booktabs}
\usepackage{tabularx}
\usepackage{pifont}
\usepackage{cuted}
\usepackage[numbers,compress]{natbib}
\usepackage{subfigure}

\usepackage{color}

\usepackage{placeins}   % 导言区
\usepackage{setspace}

\begin{document}

\title{
%Curriculum-Based Heterogeneous MADRL Based UAV Trajectory and BS Beamforming Design in Air-Ground Cooperative ISAC Netowrk
%Curriculum-Based Heterogeneous Multi-Agent Reinforcement Learning for Multi-UAV Cooperative ISAC With Joint Trajectory and Beamforming Optimization
Curriculum-Guided Heterogeneous Multi-Agent Intelligence for Multi-UAV Cooperative ISAC
}

\author{\IEEEauthorblockN{
		Kang~Yan,  \emph{Graduate~Student~Member, IEEE},
		Luping~Xiang,  \emph{Senior~Member, IEEE},
            Kang~Zheng,
            Jienan~Chen,  \emph{Senior~Member, IEEE},
            Jun~Liu,
            Qiang~Liu,  \emph{Senior~Member, IEEE},
		and~Kun~Yang}, \emph{Fellow, IEEE}\\

	\thanks{Kang~Yan and Qiang~Liu are with the School of Information and Communication Engineering, University of Electronic Science and Technology of China, Chengdu 611731, China, email: kangyan@std.uestc.edu.cn, liuqiang@uestc.edu.cn.}
	\thanks{Luping~Xiang and Kun~Yang are with the State Key Laboratory of Novel Software Technology, Nanjing University, Nanjing, China, and the School of Intelligent Software and Engineering, Nanjing University (Suzhou Campus), Suzhou, China, email: luping.xiang@nju.edu.cn, kunyang@nju.edu.cn. \textit{(Corresponding author: Luping~Xiang.)}}
    \thanks{Kang~Zheng is with the National Mobile Communications Research Laboratory, Southeast University, Nanjing 210096, China. He is also with the China Mobile Zijin Innovation Institute, Nanjing 211899, China, e-mail: zhengkang@js.chinamobile.com.}
    \thanks{Jienan~Chen is with the National Key Laboratory of Wireless Communications, University of Electronic Science and Technology of China, Chengdu, 611731, China, email: Jesson.chen@outlook.com.}
    \thanks{Jun~Liu is with the China Mobile Zijin Innovation Institute, Nanjing 211899, China, e-mail: liujun1@js.chinamobile.com.}
}
\maketitle

\begin{abstract}
	%In this paper, an air-ground cooperative integrated sensing and communication (ISAC) system is proposed, where the BS provides communication services for the unmanned aerial vehicles (UAVs) and transmits sensing signals to a sensed target, while the UAVs and BS jointly track a target by utilizing fused sensing data. We formulate a posterior Cramér-Rao bound (PCRB) minimizing optimization problem with the constraint of communication by jointly optimizing the trajectories of the multi-UAVs and the beamforming design of the BS. To solve this problem, we propose a curriculum-based heterogeneous-agent proximal policy optimization algorithm (C-HAPPO). Specifically, curriculum learning is first adopted to decompose the optimization problem into multiple tasks of increasing difficulty, and train from simple to complex tasks to ensure the convergence of the algorithm. Then, two decomposition methods, Kronecker decomposition and QR decomposition, are used to address the curse of action dimensionality. Simulation results demonstrate that our approach achieves at least 50\% improvement in sensing performance, outperforming benchmark algorithms in terms of convergence performance and target tracking accuracy.

    Seamlessly unifying communication and sensing, sixth-generation (6G) networks are poised to transform into intelligent platforms with high spectral–energy efficiency and real-time environmental awareness. In the low-altitude economy, unmanned aerial vehicles (UAVs) enable air–ground integrated sensing and communication (ISAC) for applications such as logistics and inspection, yet most studies focus on single-UAV or homogeneous-agent designs. In contrast, this paper proposes a multi-UAV cooperative ISAC system that enables heterogeneous-agent collaboration between multiple UAVs and a ground base station (BS) for joint target sensing, tracking, and communication. The system is formulated as a posterior Cramér–Rao bound (PCRB) minimization problem under communication performance constraints, utilizing joint trajectory–beamforming optimization. To tackle the NP-hard nature of this problem, we design a curriculum-based heterogeneous-agent proximal policy optimization (C-HAPPO) algorithm, where curriculum learning guides progressive policy refinement and Kronecker/QR decomposition mitigates action dimensionality. Simulation results show that the proposed approach achieves more than a 30\% improvement in sensing performance, faster convergence, and higher tracking accuracy than existing baselines, demonstrating its scalability and effectiveness for complex multi-UAV ISAC scenarios.
\end{abstract}


\section{Introduction}

% \subsection{Background}
\IEEEPARstart{I}{n} sixth-generation (6G) mobile communication systems, the paradigm will evolve from “connecting everyone” to “intelligently connecting everything” \cite{wang2023road,na2024operator}. Future networks will not only support ubiquitous connectivity but also acquire the capability to sense the environment and perform real-time inference, thus advancing toward integrated sensing and communication (ISAC) \cite{luo2025isac,luo2025bedrock,peng11165352,zhou11357472}. Unlike fifth-generation (5G) systems, which prioritize high bandwidth and low latency, 6G is envisioned to significantly enhance spectral and energy efficiency while expanding service capabilities by sharing spectrum and hardware for simultaneous data transmission and environmental sensing. This integration lays the technological foundation for transforming communication infrastructures into intelligent sensing platforms \cite{kaushik2024toward}.

Concurrently, the low-altitude economy has emerged as a new engine for regional innovation and industrial upgrading \cite{wang2025toward}. In this context, secure, reliable, and efficient communication networks are indispensable to enabling key applications such as logistics, aerial inspection, emergency response, and urban air mobility. Unmanned aerial vehicles (UAVs), with their high maneuverability, flexible deployment, broad coverage, and cost-effectiveness, can establish integrated air–ground communication infrastructures to support these applications \cite{javed2024state}. As a result, UAVs are becoming critical components of the low-altitude communication architecture, making the investigation of their communication and sensing capabilities both theoretically valuable and practically impactful.

Building upon these advances, ISAC-enabled UAV platforms offer a promising technological pathway for low-altitude economy applications \cite{jiang2025integrated,song2025overview}. By fusing wireless communication with environmental sensing, UAVs can maintain robust air–ground connectivity while providing real-time airspace monitoring, target tracking, and dynamic resource scheduling. This integration offers multidimensional information support for safe operation and intelligent management of the low-altitude economy. Owing to their high mobility, three-dimensional coverage, and mission reconfigurability, UAVs present significant opportunities for advancing ISAC research, warranting in-depth and systematic investigation.

A growing body of literature has explored UAV-based ISAC systems \cite{lyu2022joint,deng2024joint,meng2022throughput,jing2024isac,zhou2025beamforming}. For example, Lyu et al. \cite{lyu2022joint} designed a UAV as an ISAC access point to provide communication to ground users while sensing a target, jointly optimizing UAV trajectory and beamforming.
Likewise, Deng et al. \cite{deng2024joint} investigated a covert ISAC system optimizing UAV trajectory and beamforming to maximize the achievable covert communication rate for legitimate users.
Meng et al. \cite{meng2022throughput} proposed a periodic ISAC mechanism optimizing UAV trajectory, sensing time allocation, and beamforming to maximize communication throughput while ensuring sensing performance, though sensing accuracy was not explicitly quantified.
Jing et al. \cite{jing2024isac} derived a Cramér–Rao bound (CRB) to evaluate sensing accuracy, and a joint optimization framework was developed to minimize the CRB while maximizing communication throughput through UAV trajectory, bandwidth allocation, and target estimation optimization.
Similarly, Zhou et al. \cite{zhou2025beamforming} derived a CRB to evaluate the location of the loaded users and proposed a multi-UAV optimization to maximize the total achievable communication rate.
However, these studies restrict ISAC to UAV-only deployments without collaboration with terrestrial infrastructure.

Collaborative UAV–ground infrastructure frameworks have been proposed to improve ISAC coordination \cite{diaz2023sensing,liu2024uav,khalili2024efficient,wang2024isac}.
For instance, Diaz-Vilor et al. \cite{diaz2023sensing} considered a multi-cell network, where a UAV sensed and forwarded the sensing data from different events to the BS to minimize the amount of energy required.
Liu et al. \cite{liu2024uav} investigated a UAV-assisted ISAC system wherein the UAV provides Internet of Things (IoT) communication services while relaying sensing data to a data center via optimized trajectory and resource allocation. Similarly, Khalili et al. \cite{khalili2024efficient} considered a UAV that communicates with ground users and senses potential targets, forwarding the sensing data to a base station (BS). In these designs, however, UAVs primarily serve as relays for sensing data collection without addressing the operational objectives of sensing itself. In contrast, Wang et al. \cite{wang2024isac} proposed a cellular-connected UAV–BS collaboration scheme for target UAV detection, employing successive convex approximation to jointly optimize UAV trajectory and BS beamforming. Nevertheless, prior work remains limited to single-UAV–BS collaboration.

To the best of our knowledge, no existing work has studied ISAC systems involving multiple UAVs collaboratively operating with ground-based infrastructure. As the number of UAVs increases, the beamforming design complexity grows rapidly, leading to the curse of dimensionality and rendering traditional centralized convex optimization approaches intractable due to their NP-hardness. Fortunately, such multi-agent ISAC systems can be naturally modeled as Markov decision processes (MDPs) \cite{garcia2013markov}, enabling solutions via multi-agent deep reinforcement learning (MADRL). In this formulation, UAVs and the BS are modeled as autonomous agents that learn optimal strategies through interactions with the environment.

MADRL has shown significant promise in solving complex optimization problems in dynamic UAV networks. Before being applied to ISAC, learning-based approaches have demonstrated robust performance in UAV tasks. For instance, deep reinforcement learning (DRL) has been successfully employed for joint radio resource and beamforming design in cellular-connected UAVs \cite{li2023radio}, as well as for stable path planning using quantum-inspired experience replay \cite{li2022path}. Progressing towards decentralized coordination, recent work in \cite{li2025energy} verified the scalability of distributed multi-agent frameworks for UAV-enabled edge computing. Motivated by these successes, researchers have extended MADRL to the more challenging domain of UAV-enabled ISAC systems \cite{gao2024marl,xie2024distributed,cheng2024joint,qin2023deep,ye2025aoi}.
Gao et al. \cite{gao2024marl} developed a hybrid-reward multi-agent proximal policy optimization method for joint multi-UAV trajectory and beamforming optimization in a UAV-enabled ISAC framework.
Xie et al. \cite{xie2024distributed} designed a distributed multi-agent double deep Q network to optimize the energy saving efficiency of UAVs in the proposed device-free ISAC system.
Cheng et al. \cite{cheng2024joint} proposed a game-embedding MADRL algorithm to jointly optimize the UAV trajectory and ISAC task schedule.
Qin et al. \cite{qin2023deep} proposed a multi-agent soft actor–critic algorithm to jointly optimize UAV trajectory planning and power allocation. Ye et al. \cite{ye2025aoi} introduced a multi-agent curriculum learning framework to improve training stability in complex optimization tasks. However, these works focus on homogeneous agents, making them unsuitable for heterogeneous-agent systems involving both UAVs and BSs. To address this limitation, heterogeneous-agent proximal policy optimization (HAPPO) \cite{zhong2024heterogeneous} has been introduced, enabling agents with distinct action spaces to coordinate effectively in complex environments.
\begin{table*}[!t]
    \centering
    \caption{Contrasting our contributions with state-of-the-art}
    \resizebox{6in}{!}{
    \begin{tabular}{l|c|c|c|c|c|c|c|c|c}
        \hline  
		\text { Contributions } 
		& \text{this work} 
		& {\cite{lyu2022joint,deng2024joint,meng2022throughput}} 
		& {\cite{jing2024isac}} 
		& {\cite{zhou2025beamforming}} 
		& {\cite{liu2024uav}} 
		& {\cite{khalili2024efficient}} 
		& {\cite{wang2024isac}} 
		& {\cite{gao2024marl}} 
		& {\cite{xie2024distributed,cheng2024joint,qin2023deep}} 
		\\
        \hline 

        \hline 
		\text { Multi-UAVs } 
		&{\ding{52}} 
		& {		   } & {         } & {\ding{51}}& {         }& {        }& {         }& {\ding{51}}& {\ding{51}}\\
        \hline 
		\text { Trajectory design }
		&{\ding{52}} 
		& {         } & {\ding{51}} & {\ding{51}}& {\ding{51}}& {\ding{51}}& {\ding{51}}& {\ding{51}}& {\ding{51}}\\
        \hline
		\text { Beamforming design }
		&{\ding{52}} 
		& {\ding{51}} & {         } & {\ding{51}}& {         }& {\ding{51}}& {\ding{51}}& {\ding{51}}& {         }\\
        \hline
		\text { BS and UAVs collaborative} 
		&{\ding{52}} 
		& {         } & {\ding{51}} & {         }& {         }& {\ding{51}}& {\ding{51}}& {         }& {         }\\
        \hline 
		\text { Target Tracking }
		&{\ding{52}} 
		& {\ding{51}} & {         } & {\ding{51}}& {         }& {         }& {\ding{51}}& {         }& {         }\\
        \hline 
		\text { MADRL }
		&{\ding{52}} 
		& {         } & {         } & {         }& {         }& {         }& {         }& {\ding{51}}& {\ding{51}}\\
        \hline
		% \text { Curriculum Learning }
		% &{\ding{52}} 
		% & {         } & {         } & {         } & {         } & {         }& {         }& {         }& {         }& {         }& {         }& {\ding{51}}& {\ding{51}}\\
        % \hline 
		% \text { HARL }
		% &{\ding{52}} 
		% & {         } & {         } & {         } & {         } & {         }& {         }& {         }& {         }& {         }& {         }& {\ding{51}}& {\ding{51}}\\
        % \hline
    \end{tabular}
	}
    \label{Contributions}
\end{table*}

Motivated by these gaps, this paper studies a multi-UAV cooperative ISAC system in which multiple UAVs collaborate with a ground BS to jointly sense and track a target. A detailed comparison between our contributions and representative state-of-the-art works is presented in Table \ref{Contributions}, highlighting the distinctive features of this study, including the consideration of multi-UAV coordination, joint trajectory–beamforming design, UAV–BS collaborative sensing, target tracking, and the adoption of a MADRL paradigm. The primary contributions of this paper are summarized as follows:
\begin{itemize}
\item We propose an ISAC system model comprising multiple UAVs and a BS, where the BS provides communication services to the UAVs, and both the BS and UAVs jointly track a target using fused sensing data. Based on this model, we formulate a PCRB minimization problem under multi-UAV communication constraints, jointly optimizing UAV trajectories and BS beamforming.
\item To solve the above NP-hard optimization problem, we develop a curriculum-based heterogeneous-agent proximal policy optimization (C-HAPPO) algorithm. The framework integrates curriculum learning to decompose the optimization into progressively complex tasks, enabling the BS agent to learn effective beamforming strategies. Furthermore, Kronecker and QR decomposition techniques are applied to mitigate the curse of action dimensionality.
\item We carry out numerical simulations to evaluate the C-HAPPO performance of the proposed ISAC system, which demonstrate that the proposed method achieves at least a 30\% improvement in sensing performance over the original HAPPO benchmark and outperforms baseline algorithms in terms of convergence speed and target tracking accuracy. These findings highlight the scalability and effectiveness of the proposed approach in complex multi-UAV ISAC scenarios.
\end{itemize}

The remainder of this paper is organized as follows. Section II describes the system model. Section III presents the target tracking framework and problem formulation. Section IV details the proposed C-HAPPO algorithm. Section V reports simulation results and performance analysis. Section VI concludes the paper.

Notations: Bold uppercase and lowercase letters denote matrices and column vectors, respectively. $||\cdot||$ denotes the Euclidean norm, $|\cdot|$ denotes the absolute value, $(\cdot)^T$ and $(\cdot)^H$ denote the transpose and the conjugate transpose, respectively, and $\otimes$ denotes the Kronecker product.

\section{System Model}
\begin{figure}
		\centering
		\includegraphics[width=3.2in]{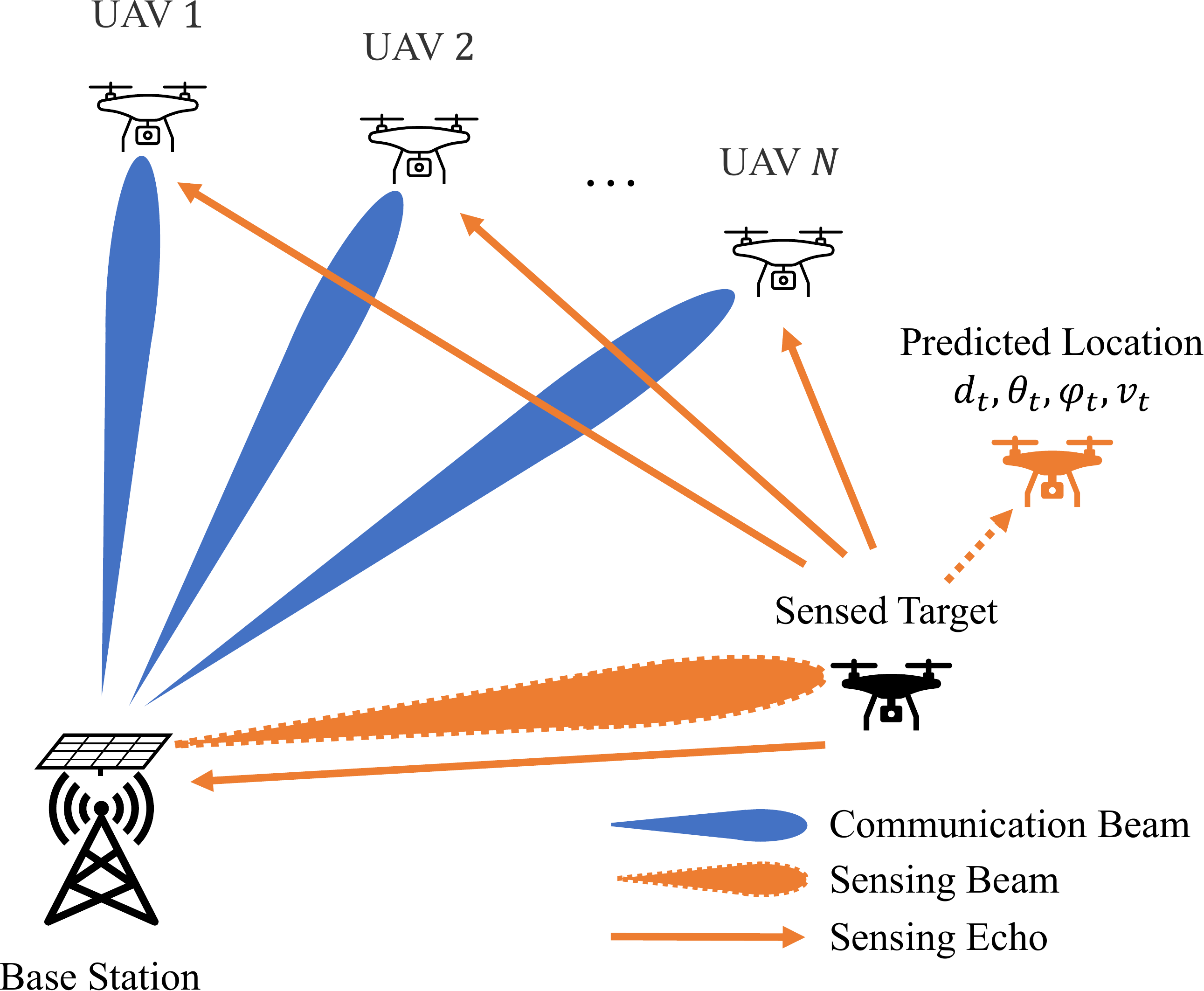}
		\caption{The proposed air-to-ground ISAC system model.}
		\label{SystemModel}
\end{figure}

As illustrated in Fig.~\ref{SystemModel}, we consider an air-to-ground ISAC architecture comprising a ground BS, $N$ cooperative UAVs, and an aerial sensed target. The BS is equipped with a uniform planar array (UPA) of $M=M_x \times M_y$ transmitting antennas and a single receiving antenna, while each UAV is equipped with a single receiving antenna. The target’s location is denoted by $\boldsymbol{\xi}_t$, and the position of the $n$-th UAV is represented by $\boldsymbol{u}_{n,t}$, where $t$ denotes the time slot. For notational consistency, the BS position is denoted as $\boldsymbol{u}_{\mathrm{bs},t}$. All UAVs are assumed to be equipped with high-accuracy global positioning system (GPS) receivers, enabling precise self-localization.

In each time slot, the BS simultaneously transmits communication signals to the UAVs and a sensing signal toward the target. The reflected sensing signal is received synchronously by the BS as well as all UAVs, facilitating cooperative estimation of the target’s position and velocity. The BS performs transmit beamforming to enhance sensing and communication performance, while each UAV optimizes its three-dimensional trajectory to improve the sensing echo's signal-to-interference-plus-noise ratio (SINR), thereby increasing sensing accuracy.

For analytical tractability, we adopt a frequency-division duplexing (FDD) framework, in which the downlink (BS-to-UAV) and uplink (UAV-to-BS) channels operate over orthogonal frequency bands. Under this assumption, the uplink does not interfere with the downlink and is therefore excluded from the system-level analysis.
Meanwhile, we assume that the locations of all UAVs can be aggregated to the BS through the uplink channel and shared with other UAVs through the downlink channel.

The rest of this section discusses the BS part and the UAV part in turn.

\subsection{Base Station Signal Model}
The BS serves a dual role by transmitting the ISAC waveform and receiving the corresponding echoes reflected from the sensed target. The BS's transmitted signal at time slot $t$ is given by
\begin{equation}
\label{BSTransmittedSignal}
\begin{aligned}
\boldsymbol{x(t)} = \sum_{n=1}^{N}\boldsymbol{w}_{n,t}s_n(t) + \boldsymbol{w}_{0,t}s_{0}(t),
\end{aligned}
\end{equation}
where $\boldsymbol{w}_{n,t}$ and $\boldsymbol{w}_{0,t}$ denote the beamforming vectors of the $n$-th UAV and of the sensed target at time slot $t$, respectively, while $s_n(t)$ and $s_{0}(t)$ are the corresponding unitary-power transmitted symbols.

Given the unobstructed propagation path between the BS and the target, the sensing channel is modeled as a Rician fading channel. The sensing channel gain between the BS and the target is
\begin{equation}
\label{BSSensingChannelGain}
\begin{aligned}
\beta_{{u}_{\mathrm{bs},t}, {\xi}_t}^{\mathrm{s}} &=\left(\sqrt{\frac{\Bbbk}{\Bbbk+1}} + \sqrt{\frac{1}{\Bbbk+1}}g\right)\sqrt{\frac{\beta_{0}^{\mathrm{s}}}{|| \boldsymbol{u}_{\mathrm{bs},t}, \boldsymbol{\xi}_t || ^{4}}},
\end{aligned}
\end{equation}
where $\Bbbk$ denotes the Rician $K$-factor, $g\sim \mathcal{CN}(0,1)$ is a complex Gaussian random variable, and $\beta^s_{0}$ denotes the reference sensing channel gain between the BS and the target at 1 m.

The received signal at the BS is given by
\begin{equation}
\label{BSReceiveSignal}
\begin{aligned}
y_{\mathrm{bs}}(t) = \alpha_0 \beta^{\mathrm{s}}_{{u}_{\mathrm{bs},t},{{\xi}_t}} \boldsymbol{a}^H_{{\xi}_t} \sum_{k=0}^{N} \boldsymbol{w}_{k,t}s_k(t) + \sum_{k=1}^{N}n_{k} +n_0,
\end{aligned}
\end{equation}
where $\alpha_0$ is the radar cross section (RCS) of the target, $n_k$ is the echo noise reflected by the $k$-th UAV, and $n_0 \sim \mathcal{CN}(0,\sigma_0^2)$ denotes additive white Gaussian noise (AWGN). The transmit steering vector $\boldsymbol{a}_{\boldsymbol{\xi}_t}$ is expressed as
\begin{equation}
\label{SteeringVector}
\begin{aligned}
\begin{aligned}
\boldsymbol{a}^{H}_{\xi_t}
= & {\left[1, \cdots, e^{\frac{-j 2 \pi d_{x}\left(M_{x}-1\right)  \psi\left(\boldsymbol{u}_{\mathrm{b s},t},\boldsymbol{\xi}_t\right)}{\lambda}}\right] } \\
& \otimes\left[1, \cdots, e^{\frac{-j 2 \pi d_{y}\left(M_{y}-1\right) \phi\left(\boldsymbol{u}_{\mathrm{b s},t}, \boldsymbol{\xi}_t\right)}{\lambda}}\right],
\end{aligned}
\end{aligned}
\end{equation}
where $d_x$ and $d_y$ are the inter-element spacings along the $x$- and $y$-axes, $M_x$ and $M_y$ denote the corresponding antenna counts, $\psi\left(\boldsymbol{u}_{\mathrm{b s},t}, \boldsymbol{\xi}_t\right)$ and $\phi\left(\boldsymbol{u}_{\mathrm{b s},t}, \boldsymbol{\xi}_t\right)$ are the direction cosines along the $x$- and $y$-axes, and $\lambda$ represents the carrier wavelength.

Assuming knowledge of the UAV positions via bidirectional BS–UAV links, echo noise terms $n_k$ can be suppressed through estimation and cancellation techniques \cite{li2017joint,li2018optimal}. In practice, cancellation is imperfect and leaves residual interference. Hence, the received signals can be reformulated as
\begin{equation}
\label{BSReceiveSignalReformulated}
\begin{aligned}
y_{\mathrm{bs}}(t) = \alpha_0 \beta^{\mathrm{s}}_{{u}_{\mathrm{bs},t},{{\xi}_t}} \boldsymbol{a}^H_{{\xi}_t} \sum_{k=0}^{N} \boldsymbol{w}_{k,t}s_k(t) + I_{u_{\text{bs}}}(t) +n_0,
\end{aligned}
\end{equation}
where $I_{u_{\text{bs}}}(t) =I_{u_{\text{bs},\text{res}}}(t)+ I_{u_{\text{bs},\text{mp}}}(t) \sim \mathcal{CN}(0,\sigma^2_{u_{\text{bs}}})$ includes residual cancellation noise term $I_{u_{\text{bs},\text{res}}}(t)$ and unresolved multipath interference noise term $I_{u_{\text{bs},\text{mp}}}(t)$.

Consequently, the sensing SINR at the BS is
\begin{equation}
\label{BSSensingSINR}
\begin{aligned}
\mathrm {SINR}^{\mathrm{s}}_{u_{bs}}(t)   = \frac
{\left|\alpha \beta^s_{u_{\mathrm{bs},t},{\xi_t}} \boldsymbol{a}^H_{\xi_t} \boldsymbol{w}_{0,t}\right|^2}
{\sum_{k=1}^{N}\left|\alpha \beta^s_{u_{\mathrm{bs},t},{\xi_t}} \boldsymbol{a}^H_{\xi_t} \boldsymbol{w}_{k,t}\right|^2+
\sigma^2_{u_{\text{bs}}}+
\sigma^2_{0}}.
\end{aligned}
\end{equation}

\subsection{UAV Signal and Sensing Model}
Unlike the BS, each UAV operates solely as a receiver, capturing both downlink communication signals and sensing echoes. The sensing channel gain from the BS to the $n$-th UAV via the target is
\begin{equation}
\label{UAVSensingChannelGain}
\begin{aligned}
\beta_{u_{n,t}, \xi_t}^{\mathrm{s}} &=\left(\sqrt{\frac{\Bbbk}{\Bbbk+1}} + \sqrt{\frac{1}{\Bbbk+1}}g\right)\\
&  \quad \quad\sqrt{\frac{\beta_{0}^{\mathrm{s}}}{|| \boldsymbol{u}_{\mathrm{bs},t}, \boldsymbol{\xi}_t || ^{2} || \boldsymbol{u}_{n,t}, \boldsymbol{\xi}_t || ^{2}}},
\end{aligned}
\end{equation}
and the communication channel gain between the BS and the $n$-th UAV is
\begin{equation}
\label{CommChannelGain}
\begin{aligned}
\beta^{\mathrm{c}}_{u_{\mathrm{bs},t},{u_{n,t}}} = \left(\sqrt{\frac{\Bbbk}{\Bbbk+1}} + \sqrt{\frac{1}{\Bbbk+1}}g\right)\sqrt{\frac{\beta^c_0}{||\boldsymbol{u}_{\mathrm{bs},t},\boldsymbol{u}_{n,t} || ^2}},
\end{aligned}
\end{equation}
where $\beta^{\mathrm{c}}_0$ is the reference communication channel gain between the BS and the UAVs at 1 m.

The received signal $y_{u_n}(t)$ at the $n$-th UAV is expressed in \eqref{UAVReceiveSignal}, where $\alpha_1$ is the RCS of the cooperative UAVs.
%After eliminating the common echo noise terms $n_k$ (cf. \eqref{BSReceiveSignal}), the downlink communication SINR $\mathrm {SINR}^c_{u_n} (t)$ at UAV $n$ is expressed as \eqref{UAVCommSINR}.
The downlink communication SINR $\mathrm {SINR}^c_{u_n} (t)$ at UAV $n$ is expressed as \eqref{UAVCommSINR}.
\begin{figure*}[!t]     % [!b] = 尽量放页面底端
\centering
% \rule{\textwidth}{0.4pt}\\[-1.0ex]         % 横线
\begin{equation}
\label{UAVReceiveSignal}
\begin{aligned}
y_{u_n}(t) =  \beta^{\mathrm{c}}_{u_{\mathrm{bs},t},{u_{n,t}}} \boldsymbol{a}^H_{u_{n,t}} \sum_{k=0}^{N} \boldsymbol{w}_{k,t}s_k(t) + \alpha_0 \beta^s_{u_{n,t},\xi_t} \boldsymbol{a}^H_{\xi_t} \sum_{k=0}^{N} \boldsymbol{w}_{k,t}s_k(t) + \sum_{i=1, i\ne n}^{N} \alpha_1 \beta^s_{u_{n,t},u_{i,t}} \boldsymbol{a}^H_{u_{i,t}} \sum_{k=0}^{N} \boldsymbol{w}_{k,t}s_k(t)  +n_0
\end{aligned}.
\end{equation}

\begin{equation}
\label{UAVCommSINR}
\begin{aligned}
\mathrm {SINR}^{\mathrm{c}}_{u_n} (t)   = 
\frac{\left| \beta^{\mathrm{c}}_{u_{\mathrm{bs},t},{u_{n,t}}} \boldsymbol{a}^H_{u_{n,t}} \boldsymbol{w}_{n,t}\right|^2}
{ \sum\limits_{k=1, k\ne n}^{N}\left|\beta^{\mathrm{c}}_{u_{\mathrm{bs},t},{u_{n,t}}} \boldsymbol{a}^H_{u_{n,t}} \boldsymbol{w}_{k,t}\right|^2+
\sum\limits_{k=0}^{N}\left|\alpha_0 \beta^{\mathrm{s}}_{u_{n,t},{\xi_t}} \boldsymbol{a}^H_{\xi_t}  \boldsymbol{w}_{k,t}\right|^2 + 
\sum\limits_{i=1,i\ne n}^{N}\sum\limits_{k=0}^{N}\left|\alpha_1 \beta^{\mathrm{s}}_{u_{n,t},u_{i,t}} \boldsymbol{a}^H_{u_{i,t}}  \boldsymbol{w}_{k,t}\right|^2 +\sigma^2_0}
\end{aligned}.
\end{equation}
\rule{\textwidth}{0.4pt}
\end{figure*}

Direct computation of the sensing SINR at the UAV is severely impaired by the strong interference from downlink communication signals.
To mitigate this effect, the UAV forwards its complete received signal to the BS via the uplink channel, and the BS maintains an exact copy of the transmitted communication waveform.
Leveraging this reference, the BS can perform interference cancellation of the received communication waveform of the UAVs, thereby enabling estimation of the sensing SINR at the UAV. The received signals from UAV $u_n$ can be reformulated as
\begin{equation}
\label{UAVReceiveSignalS}
\begin{aligned}
y_{u_n}(t) =  \alpha_0 \beta^s_{u_{n,t},\xi_t} \boldsymbol{a}^H_{\xi_t} \sum_{k=0}^{N} \boldsymbol{w}_{k,t}s_k(t) + I_{u_{n}}(t)  +n_0
\end{aligned}.
\end{equation}
where $I_{u_{n}}(t) =I_{u_{n,\text{res}}}(t)+ I_{u_{n,\text{mp}}}(t) \sim \mathcal{CN}(0,\sigma^2_{u_{n}})$ is similar to $I_{u_{\text{bs}}}(t)$, including residual cancellation noise term $I_{u_{n,\text{res}}}(t)$ and unresolved multipath interference noise term $I_{u_{n,\text{mp}}}(t)$.

The resulting sensing SINR for the $n$-th UAV at time slot $t$ is given by
\begin{equation}
\label{UAVSensingSINR}
\begin{aligned}
\mathrm {SINR}^{\mathrm{s}}_{u_{n}}(t)   = \frac
{\left|\alpha_0 \beta^{\mathrm{s}}_{u_{n,t},{\xi_t}} \boldsymbol{a}^H_{\xi_t} \boldsymbol{w}_{0,t}\right|^2}
{\sum\limits_{k=1}^{N}\left|\alpha_0
\beta^{\mathrm{s}}_{u_{n,t},{\xi_t}} \boldsymbol{a}^H_{\xi_t} \boldsymbol{w}_{k,t}\right|^2+
\sigma^2_{u_{n}}+
\sigma^2_{0}}.
\end{aligned}
\end{equation}

It is worth noting that we consider a co-frequency system where the communication and sensing signals occupy the same time-frequency resources. While advanced delay-Doppler processing could potentially exploit waveform orthogonality to separate signals, we adopt a worst-case assumption where signals are treated as non-orthogonal in the delay-Doppler domain. Therefore, the sensing performance in our model is fundamentally limited by the time-frequency domain SINR derived above.

This BS-assisted sensing mechanism inherently depends on sufficient uplink reporting from the UAVs. Under communication-limited conditions, incomplete, delayed, or compressed uplink observations reduce the accuracy of interference cancellation and sensing-SINR estimation at the BS, which further degrades sensing-information fusion and leads to a larger PCRB as well as worse tracking performance.

\section{Target Tracking and Problem Formulation}
This section first outlines the fusion of sensing measurements collected by the UAVs and the BS, followed by the derivation of the tracking performance metric. It then introduces the target motion and measurement models, and finally formulates the performance optimization problem.

\subsection{Sensing Information Fusion}
During tracking, the BS and UAVs estimate the target’s radial distance $d$, polar angle $\theta$, azimuth angle $\varphi$, and radial velocity $v$ in the spherical coordinate system with respect to the BS, compactly expressed as
\begin{equation}
\label{EstimationParas}
\begin{aligned}
\boldsymbol{c}_t = \begin{bmatrix}
d_{t},& \theta_{t}, & \varphi_{t},  &v_{t}
\end{bmatrix}^T
\end{aligned}
\end{equation}

Although the mean-square error (MSE) is a standard performance metric, its closed-form expression is generally intractable in practical ISAC systems. Therefore, the Cramér–Rao bound (CRB) is adopted as a fundamental lower bound on the MSE. 
The CRBs for $d_t$, $\theta_t$, $\varphi_t$, and $v_t$ are denoted by $\sigma_{d_t}^{2}$,  $\sigma_{\theta_t}^{2}$, $\sigma_{\varphi_t} ^{2}$, and $\sigma_{v_t}^{2}$, respectively, and can be expressed in the unified form \cite{dong2022sensing}
\begin{equation}
\label{SensingCRB}
\begin{aligned}
\sigma^2_{d_t,(\cdot)}  = \frac{\kappa_{d}}{\left|\mathrm {SINR}^s_{(\cdot)}(t)\right|^2 b}, \;\;
&\sigma^2_{\theta_t,(\cdot)}  = \frac{\kappa_{\theta}}{\left|\mathrm {SINR}^s_{(\cdot)}(t)\right|^2},\\
\sigma^2_{\varphi_t,(\cdot)}  = \frac{\kappa_{\varphi}}{\left|\mathrm {SINR}^s_{(\cdot)}(t)\right|^2}, \;\;
&\sigma^2_{v_t,(\cdot)}  = \frac{\kappa_{v}}{\left|\mathrm {SINR}^s_{(\cdot)}(t)\right|^2},
\end{aligned}
\end{equation}
where $\kappa_{d}$, $\kappa_{\theta}$, $\kappa_{\varphi}$, and $\kappa_{v}$ are their respective correlation constants about waveform and geometry dependent factors \cite{liu2020radar}, and $b$ is the signal bandwidth.

As the correlation between the BS and UAV estimates is generally unknown, the BS fuses the CRBs from UAVs and BS using the Covariance Intersection (CI) method, yielding
\begin{equation}
\label{FusedCRB}
\begin{aligned}
\bar{\sigma}^2_{(\cdot)}=\left(\omega_{0} \left( \sigma^2_{(\cdot),u_{bs}}\right)^{-1} +\sum_{n =1}^{N} \omega_{n}\left( \sigma^2_{(\cdot),u_n}\right)^{-1}\right)^{-1},
\end{aligned}
\end{equation}
where $\omega_{n}$ are non-negative fusion weights satisfying
\begin{equation}
\label{FusedCRBCoefficients}
\begin{aligned}
\sum_{n=0}^{n=N} \omega_{n}=1 ,\;\;\omega_{n}\ge0.
\end{aligned}
\end{equation}

\subsection{Tracking Measurement Metric}
\subsubsection{Target Motion State Model}

Throughout the entire operating period, the target’s motion type is a priori unknown and may follow arbitrary trajectories, including uniform motion, constant acceleration, or complex maneuvers. However, directly modeling such arbitrary dynamics is computationally intractable. To ensure tractability, we approximate the target dynamics using a nearly constant velocity within each time slot \cite{blair2021industry}, where its Cartesian state vector is
\begin{equation}
\label{MotionStateVector}
\begin{aligned}
\boldsymbol{\zeta}_t = [\xi_{x,t},\xi_{y,t},\xi_{z,t},\dot{\xi}_{x,t},\dot{\xi}_{y,t},\dot{\xi}_{z,t}]^T,
\end{aligned}
\end{equation}
where the first three elements denote position and the last three denote velocity. The state evolution follows
\begin{equation}
\label{MotionStateModel}
\begin{aligned}
\boldsymbol{\zeta}_t = \mathbf{F}_{\zeta}\boldsymbol{\zeta}_{t-1} + \mathrm {\boldsymbol{\varpi}}_{t-1},
\end{aligned}
\end{equation}
with the state transition matrix
\begin{equation}
\label{StateTransferMatrix}
\begin{aligned}
\mathbf{F}_{\zeta_t} = \begin{bmatrix}
1 & T \\
0 & 1 \\
\end{bmatrix}
\otimes \boldsymbol{\mathrm{I}}_3,
\end{aligned}
\end{equation}
where $T$ is the slot duration and $ \mathrm {\boldsymbol{\varpi}}_{t-1}$ is zero-mean Gaussian process noise with covariance
\begin{equation}
\label{ProcessNoiseCovariance}
\begin{aligned}
\mathbf{\Phi}_{\zeta_t} = \begin{bmatrix}
\frac{1}{3}T^3 &  \frac{1}{2}T^2 \\
\frac{1}{2}T^2 & T \\
\end{bmatrix}
\otimes\sigma^2_{\zeta_t}  \boldsymbol{\mathrm{I}}_3,
\end{aligned}
\end{equation}
where $\sigma^2_{\zeta_t}$ is the process noise intensity, including acceleration errors and maneuvers.

\subsubsection{Target Measurement State Model}
The nonlinear measurement model of the sensed target can be written as:
\begin{equation}
\label{NonlinearMeasurementModel}
\begin{aligned}
\boldsymbol{y}_t = h(\boldsymbol{\zeta}_{t}) + \tilde{ \mathrm {\boldsymbol{\varpi}}}_{t},
\end{aligned}
\end{equation}
where $\boldsymbol{y}_t$ is specified in (\ref{EstimationParas}) and $h(\cdot)$ represents the nonlinear transformation from the Cartesian coordinate system to the spherical coordinate system, which can be written as
\begin{equation}
\label{CartesianTosphericalTransformation}
\begin{aligned}
h\left(\boldsymbol{\zeta}_{t}\right)=\left\{\begin{array}{l}
d_{t}=\sqrt{\xi_{x,t}^{2}+\xi_{y,t}^{2}+\xi_{z,t}^2}, \\
\theta_{t}=\arccos \left(\xi_{z,t} / d_{t}\right),\\
\varphi_{t}=\arctan \left(\xi_{y,t} / \xi_{x,t}\right),\\
v_{t}=\left(\dot{\xi}_{x,t} \xi_{x,t}+\dot{\xi}_{y,t} \xi_{y,t} + \dot{\xi}_{z,t}\xi_{z,t}\right) / d_{t} .\\
\end{array}\right.
\end{aligned}
\end{equation}
The measurement noise $\tilde{ \mathrm {\boldsymbol{\varpi}}}_{t}$ is modeled as zero-mean AWGN with covariance
\begin{equation}
\label{MeasurementNoise}
\begin{aligned}
\mathbf{\Psi}_{t} =\mathrm {diag} (\bar{\sigma}^2_{d_t},\bar{\sigma}^2_{\theta_t},\bar{\sigma}^2_{\varphi_t},\bar{\sigma}^2_{ v_t})
\end{aligned}
\end{equation}
where the CRB terms are given by \eqref{FusedCRB}.

\subsubsection{Tracking Performance}
The tracking performance is quantified via the PCRB. Accordingly, the first step is to compute the posterior Fisher information matrix (FIM), which is expressed as
\begin{equation}
\label{PCRBDecomposition}
\begin{aligned}
\mathbf{J} \left(\boldsymbol{\zeta}_{t}\right) = \mathbf{J}_P \left(\boldsymbol{\zeta}_{t}\right) +\mathbf{J}_D \left(\boldsymbol{\zeta}_{t}\right),
\end{aligned}
\end{equation}
where $ \mathbf{J}_P \left(\boldsymbol{\zeta}_{t}\right) $ represents the prior FIM, which is
\begin{equation}
\label{PriorFIM}
\begin{aligned}
\mathbf{J}_P \left(\boldsymbol{\zeta}_{t}\right)  = \left( 
\mathbf{\Phi}_{\zeta_t}+  \mathbf {{F}}_{\zeta_t} \mathbf{J} \left(\boldsymbol{\zeta}_{t-1}\right)^{-1} \mathbf{{F}}_{\zeta_t}^T
\right)^{-1}.
\end{aligned}
\end{equation}
$\mathbf{J}_D \left(\boldsymbol{\zeta}_{t}\right)$ denotes the data FIM, which is written as
\begin{equation}
\label{DataFIM}
\begin{aligned}
\mathbf{J}_D \left(\boldsymbol{\zeta}_{t}\right)  = \hat{\mathbf{H}}_{\zeta_t}^T \hat{\mathbf{\Psi}}_{t} \hat{\mathbf{H}}_{\zeta_t},
\end{aligned}
\end{equation}
where $\hat{\mathbf H}_{\boldsymbol{\zeta}_t}$ indicates the Jacobian of $h(\boldsymbol{\zeta}_{t})$ evaluated at the predicted target state $\hat{\boldsymbol{\zeta}}_{t}$, and $\hat{\boldsymbol{\Psi}}_{t}$ represents the measurement noise covariance matrix corresponding to $\hat{\boldsymbol{\zeta}}_{t}$.

The PCRB at time slot $t$ is the inverse of the FIM matrix $\mathbf{J} \left(\boldsymbol{\zeta}_{t}\right)$, which is given by
\begin{equation}
\label{PCRB}
\begin{aligned}
\text {PCRB}_t = \mathbf{J} \left(\boldsymbol{\zeta}_{t}\right) ^{-1}.
\end{aligned}
\end{equation}

The sensing performance can be measured using the trace of the PCRB matrix. This metric represents the aggregate variance of the position and velocity states and can be computed as
\begin{equation}
\label{TracePCRB}
\begin{aligned}
\rho_t = \mathrm{trace}\left(\text {PCRB}_t\right) .
\end{aligned}
\end{equation}

\subsection{Tracking Algorithm}
To enable high-accuracy target tracking, an extended Kalman filter (EKF) is employed, leveraging current sensing measurements to predict the target state in the subsequent time slot and applying resource allocation to refine tracking accuracy.  The steps of the EKF are as follows \cite{dong2022sensing}

\subsubsection{Motion State Prediction}
\begin{equation}
\label{MotionStatePrediction}
\begin{aligned}
\hat{\boldsymbol{\zeta}}_{t|t-1} = \mathbf{F}_{\zeta}\hat{\boldsymbol{\zeta}}_{t-1}.
\end{aligned}
\end{equation}

\subsubsection{MSE Matrix Prediction}
\begin{equation}
\label{MSEMatrixPrediction}
\begin{aligned}
\mathbf{M}_{t|t-1} = \mathbf{F}_{\zeta} \mathbf{M}_{t-1} \mathbf{F}_{\zeta}^T + \mathbf{\Phi}_{\zeta},
\end{aligned}
\end{equation}
where $\mathbf{M}_{t-1}$ is equal to the PCRB in \eqref{PCRB}.

\subsubsection{Kalman Gain Calculation}
\begin{equation}
\label{almanGainCalculation}
\begin{aligned}
\mathbf{K}_{t} = \mathbf{M}_{t|t-1} \hat{\mathbf{H}}_{\zeta_t}^T \left(\hat{\mathbf{\Psi}}_{t} + \hat{\mathbf{H}}_{\zeta_t} \mathbf{M}_{t|t-1} \hat{\mathbf{H}}_{\zeta_t}^T \right)^{-1}.
\end{aligned}
\end{equation}

\subsubsection{Motion State Tracking}
\begin{equation}
\label{MotionStateTracking}
\begin{aligned}
\hat{\boldsymbol{\zeta}}_{t}= \hat{\boldsymbol{\zeta}}_{t|t-1} + \mathbf{K}_{t} \left( \boldsymbol{y}_t - h(\hat{\boldsymbol{\zeta}}_{t|t-1})\right).
\end{aligned}
\end{equation}

\subsubsection{MSE Matrix Update}
\begin{equation}
\label{MSEMatrixUpdate}
\begin{aligned}
\mathbf{M}_t = \left( \mathbf{I} -  \mathbf{K}_{t} \hat{\mathbf{H}}_{\zeta_t} \right)\mathbf{M}_{t|t-1}.
\end{aligned}
\end{equation}

\subsection{Problem Formulation}

We jointly design the BS beamforming vectors and UAV trajectories to enhance both communication and sensing performance. 
The objective is to minimize the time-averaged PCRB of the target estimate, subject to communication quality and flight-safety constraints. The optimization is formulated as
\begin{align}
\mathbf{P}: \min _{\boldsymbol{w}_n, \boldsymbol{u}_n} & \lim_{T \to \infty} \frac{1}{T} \sum_{t=1}^{T} \rho_t  \label{OptimizationProblem}\\
\text { s.t. } 
& \mathrm {SINR}^c_{u_n} (t) \geq \gamma_{\text{min}}, \tag{\ref{OptimizationProblem}{a}} \label{OptimizationProblema}\\
& ||\Delta \boldsymbol{u}_{n,t}|| \le \Delta u_{\text{max}},\; \tag{\ref{OptimizationProblem}{b}} \label{OptimizationProblemb}\\
& ||\Delta \boldsymbol{d}_{m,n}|| > \Delta d_{\text{min}},\; m \ne n, \tag{\ref{OptimizationProblem}{c}} \label{OptimizationProblemc}\\
& \sum_{n=0}^{N}\begin{Vmatrix} \boldsymbol{w}_{n,t} \end{Vmatrix}^2 = P, \tag{\ref{OptimizationProblem}{d}} \label{OptimizationProblemd}
\end{align}
where \eqref{OptimizationProblema} enforces a minimum downlink communication quality for each UAV, \eqref{OptimizationProblemb} and \eqref{OptimizationProblemc} specify the flight safety for the UAVs, ensuring that their speed $||\Delta \boldsymbol{u}_{n,t}|| = ||\boldsymbol{u}_{n,t} - \boldsymbol{u}_{n,t-1}||$ does not exceed $\Delta u_{\text{max}}$ and requiring that the distance $ ||\Delta\boldsymbol{d}_{m,n}|| = ||\boldsymbol{u}_{m,t} - \boldsymbol{u}_{n,t}||$ between any two UAVs be at least $\Delta d_{\text{min}}$, respectively, and \eqref{OptimizationProblemd} specifies the BS per-slot transmit-power budget $P$.

The optimization problem $\mathbf{P}$ is  NP-hard due to the nonconvex coupling between beamforming and trajectory variables in both the objective and constraints, as well as temporal coupling induced by mobility. To address this challenge, a C-HAPPO algorithm is developed in the next section.

\section{MADRL Algorithm Design}
This section first casts the joint design problem into a MADRL framework, then outlines the HAPPO baseline and details the proposed C-HAPPO  algorithm, followed by a summary of the overall workflow.
% Finally, we analyze the computational complexity of the proposed C-HAPPO algorithm.
\subsection{MADRL Framework}
MADRL considers multiple agents that interact with a shared environment simultaneously. Each agent learns a decision policy from locally observed information and scalar rewards, and the actions of all agents jointly shape the environment dynamics.
To apply MADRL to the MDP-formulated problem, we specify the MDP components as follows.

\subsubsection{Agent}
Each UAV and the BS is modeled as an agent. Every agent selects its action based on its current observation.

\subsubsection{State}
After all agents act, the environment transitions to a new state, which is used to evaluate policy quality during training. In the proposed system, the state comprises: (i) a \emph{communication state} (the UAVs’ SINRs and their locations), and (ii) a \emph{sensing state} (the trace of the PCRB and the target motion state). The state at slot $t$ is
\begin{equation}
\label{State}
\begin{aligned}
\boldsymbol{s}_t = \left\{ \boldsymbol{\mathrm {SINR}}^c(t), \boldsymbol{u}_t, \rho_t,  \boldsymbol{\zeta}_t \right\},
\end{aligned}
\end{equation}
where $\boldsymbol{\mathrm {SINR}}^c(t) = \left\{ \mathrm {SINR}_1^c(t),\dots,\mathrm {SINR}_N^c(t)\right\}$.

\subsubsection{Observation}
In practice, agents cannot access the full state and instead rely on local sensing and inter-agent communications.
Under the assumption of the uplink and downlink channels, the communication state of all UAVs is directly observed by each agent.
For the sensing state, the PCRB is derived from received signals, while the target state is predicted via the EKF. The observation used for control is
\begin{equation}
\label{Observation}
\begin{aligned}
\boldsymbol{o}_t = \left\{ \boldsymbol{\mathrm {SINR}}^c(t), \boldsymbol{u}_t, \rho_t, \hat{ \boldsymbol{\zeta}}_t \right\}.
\end{aligned}
\end{equation}

\subsubsection{Action}
At each slot, each agent selects an action conditioned on its observation. UAV agents choose the next-slot waypoint, while the BS agent chooses next-slot beamformers for all UAVs and the sensed target:
\begin{equation}
\label{Action}
\begin{aligned}
\boldsymbol{a}_{u_n,t} &= \left\{\boldsymbol{u}_{n,{t+1}}\right\},\\
\boldsymbol{a}_{bs,t} &= \left\{\boldsymbol{w}_{1,{t+1}},\dots,\boldsymbol{w}_{N,{t+1}},\boldsymbol{w}_{0,{t+1}}\right\}.
\end{aligned}
\end{equation}

When the UAV agent generates the waypoint for the next time slot, a mapping function
$\|\Delta \mathbf{u}_{n,t+1}\| \le \Delta u_{\max}$ is applied to ensure that the action strictly satisfies the maximum speed constraint defined in \eqref{OptimizationProblemb}. Moreover, a collision-avoidance projection is further imposed on the generated waypoints to guarantee the minimum inter-UAV separation constraint in \eqref{OptimizationProblemc}.
Similarly, regarding the base station action, the BS actor network outputs an unnormalized beamforming matrix, which is subsequently normalized by
$\mathbf{a}_{bs,t}=\frac{\sqrt{P}\mathbf{a}_{bs,t}}{\|\mathbf{a}_{bs,t}\|^2}$
to enforce the total power constraint defined in \eqref{OptimizationProblemd}.

\subsubsection{Reward}

Within the MADRL framework, agents learn policies that maximize cumulative reward from the environment. For our problem, the reward function is constructed using the Lagrange multiplier method. 
Since constraints \eqref{OptimizationProblemb}, \eqref{OptimizationProblemc} and \eqref{OptimizationProblemd} are satisfied by the action design, the reward function comprises two terms: one for minimizing the average PCRB, and one for enforcing the communication constraint \eqref{OptimizationProblema}.

For minimizing the average PCRB, the PCRB minimum reward component $r^s(t)$ can be formulated as
\begin{equation}
\label{RewardPCRB}
\begin{aligned}
r^s(t) = -\rho_t.
\end{aligned}
\end{equation}

To satisfy the communication constraint, the corresponding reward component $r^c_n(t)$ for the $n$-th UAV is defined as follows
\begin{equation}
\label{RewardComm}
\begin{aligned}
r^c_n(t) = \left\{\begin{array}{l}
	\mathrm {SINR}^c_{u_n} (t) - \gamma_{\text{min}}, \;\;  \mathrm {if} \;\; \mathrm {SINR}^c_{u_n} (t) < \gamma_{\text{min}},  \\
	0,\;\; \mathrm {otherwise}.\\ 
\end{array}\right.
\end{aligned}
\end{equation}

% Additionally, to enforce the safety distance constraint, the corresponding reward term $r^d_{m,n}(t)$ between the $m$-th UAV and the $n$-th UAV is given by
% \begin{equation}
% \label{RewardDistance}
% \begin{aligned}
% r^d_{m,n}(t) = \left\{\begin{array}{l}
% 	\Delta d_{m,n} - \Delta d_{\text{min}}, \;\;  \mathrm {if} \;\;\Delta d_{m,n} < \Delta d_{\text{min}},  \\
% 	0,\;\; \mathrm {otherwise}.\\ 
% \end{array}\right.
% \end{aligned}
% \end{equation}

The overall per-slot reward is
\begin{equation}
\label{Reward}
\begin{aligned}
r(t) = \lambda^s r^s(t) + \lambda^c \sum_{n=1}^{N} r^c_n(t)
\end{aligned},
\end{equation}
where $\lambda^s$ and $\lambda^c$ denote the weighting coefficients for the PCRB term and the communication constraint term, respectively.

\subsection{HAPPO Algorithm}
The proposed C-HAPPO algorithm extends the HAPPO algorithm \cite{zhong2024heterogeneous} to address the proposed optimization problem, which extends multi-agent proximal policy optimization (MAPPO) to heterogeneous multi-agent systems by introducing a sequential update mechanism. HAPPO operates under a centralized training decentralized execution (CTDE) actor-critic framework, where a critic network ingests the global state from the environment and outputs a value estimate $V$ for policy evaluation, while each actor network takes its local observation and outputs the corresponding action.

In HAPPO, the critic network parameters $\delta_k$ are optimized by minimizing the temporal difference (TD) error, which is defined as
\begin{equation}
\label{CriticUpdate}
\begin{aligned}
\delta_{k+1}=\arg \min _{\delta_k} \frac{1}{B T} \sum_{b=1}^{B} \sum_{t=0}^{T}\left(V_{\delta_k}\left(s_{t}\right)-{R}_{t}\right)^{2},
\end{aligned}
\end{equation}
where $B$ indicates the mini-batch size of experience buffer for training and ${R}_{t}$ denotes the discount return in the time slot $t$, given by

\begin{equation}
\label{DiscountReturn}
\begin{aligned}
{R}_{t} = \sum_{\tau=0}^{\infty } \gamma^{\tau} r\left( t+\tau+1 \mid s_t\right),
\end{aligned}
\end{equation}
where $\gamma$ represents the discount factor.

\begin{figure*}[!t]     % [!b] = 尽量放页面底端
\centering
% \rule{\textwidth}{0.4pt}\\[-1.0ex]         % 横线
\begin{equation}
\label{ActorUpdate}
\begin{aligned}
J (\vartheta_{k+1}^{n}) = \frac{1}{B T} \sum_{b=1}^{B} \sum_{t=0}^{T} \min \left(\frac{\pi_{\vartheta^n}^{n}\left(a_{t}^{n} \mid o_{t}^{n}\right)}{\pi_{\vartheta_{k}^{n}}^{n}\left(a_{t}^{n} \mid o_{t}^{n}\right)} M^{n}\left(s_{t}, \boldsymbol{a}_{t}\right), \operatorname{clip}\left(\frac{\pi_{\vartheta^{n}}^{n}\left(a_{t}^{n} \mid o_{t}^{n}\right)}{\pi^n_{\vartheta_{k}^{n}}\left(a_{t}^{n} \mid o_{t}^{n}\right)}, 1 \pm \epsilon\right) M^{n}\left(s_{t}, \boldsymbol{a}_{t}\right)\right)
\end{aligned}.
\end{equation}
%   {\footnotesize 公式(\ref{eq:big2})：说明文字。}\\[0.2ex] 
\rule{\textwidth}{0.4pt}
\end{figure*}

For the actor network of the $n$-th agent, the network parameters are updated by (\ref{ActorUpdate}), where $\operatorname{clip}(\mathcal{R},1\pm\epsilon)$ denotes truncating the ratio $\mathcal{R}$ onto the interval $[1-\epsilon,\,1+\epsilon]$, which can be expressed as
\begin{equation}
\label{Truncating}
\begin{aligned}
\operatorname{clip}(\mathcal{R},1\pm\epsilon)=\min\{\max(\mathcal{R},\,1-\epsilon),\,1+\epsilon\},
\end{aligned}
\end{equation}
where $\epsilon$ denotes the clipping threshold. The iterative equation $M^{n+1}\left(s_{}, \boldsymbol{a}_{}\right)$ is defined as
\begin{equation}
\label{IterativeEquation}
\begin{aligned}
M^{n+1} \left(s_{}, \boldsymbol{a}_{}\right) &=\frac{\pi_{\vartheta^{n}_{k+1}}^{n}\left(a_{}^{n} \mid \boldsymbol{o}_{}^{n}\right)}{\pi_{\vartheta^{n}_{k}}^{n}\left(a_{}^{n} \mid \boldsymbol{o}_{}^{n}\right)}M^{n}\left(s_{}, \boldsymbol{a}_{}\right), \\ 
M^{{1}}\left(s_{}, \boldsymbol{a}_{}\right) &=\hat{A}(s,\boldsymbol{a}),
\end{aligned}
\end{equation}
where $\hat{A}(s,\boldsymbol{a})$ indicates the advantage function calculated using the $V$ value based on the generalized advantage estimation (GAE) method \cite{schulman2015high}.

\subsection{C-HAPPO Algorithm Design}
In the proposed C-HAPPO algorithm, which builds on the HAPPO framework, we employ curriculum learning to organize training tasks by difficulty level, in order to enhance convergence. We also introduce Kronecker decomposition and QR decomposition to address the curse of dimensionality.

\subsubsection{Curriculum Learning}
Curriculum learning organizes training tasks by difficulty to facilitate mastery of complex objectives. Inspired by human pedagogy, the algorithm first tackles simple tasks, then progresses through increasingly challenging tasks until the target task is completed. In C‑HAPPO, jointly satisfying the communication SINR and sensing PCRB requirements via beamforming is difficult, due to the risk of convergence in a suboptimum. To mitigate this, we grade tasks using the communication SINR threshold and the sensing CRB coefficients as parameters. These grading parameters are defined as follows
\begin{equation}
\label{GradingParameters}
\begin{aligned}
\mathfrak{R}  = \left\{ \gamma_{\text{min}}, \kappa_{d}, \kappa_{\theta}, \kappa_{\varphi}, \kappa_{v}\right\}.
\end{aligned}
\end{equation}

During curriculum learning, initial tasks use a reduced grading parameter set $\mathfrak{R}_1$ to enable the BS agent to identify the optimal beamforming direction. After the cumulative reward exceeds a predefined threshold ${R}_{\text{min}}$, the grading parameters in $\mathfrak{R}$ are increased to raise the task difficulty. The $l$-th task grading parameters $\mathfrak{R}_l$ can be computed as
\begin{equation}
\label{GradingParametersUpate}
\begin{aligned}
\mathfrak{R}_l  = \frac{l\left( \mathfrak{R}_{\text{L}} - \mathfrak{R}_1 \right)}{L} + \mathfrak{R}_1,
\end{aligned}
\end{equation}
where $L$ indicates the total number of the curriculum learning tasks and $\mathfrak{R}_{\text{L}}$ represents the final grading parameters, corresponding to the actual parameters in the proposed system model.

After updating the grading parameters, the agents reload their models to retain the learned network parameters.
\begin{figure*}
\centering
\includegraphics[width=6in]{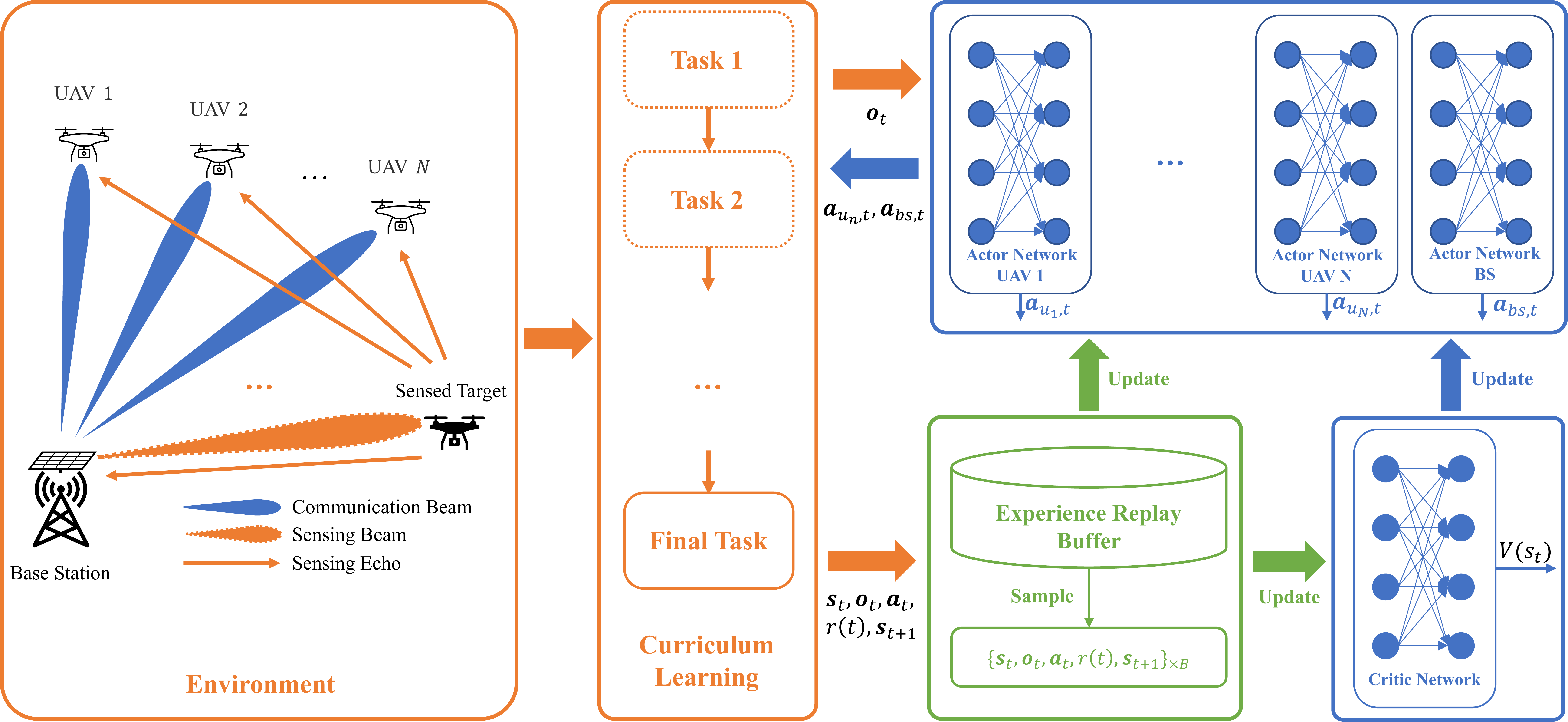}
\caption{Operation process of the C-HAPPO algorithm for the proposed system model.}
\label{AlgorithmFramework}
\end{figure*}
\subsubsection{Kronecker Decomposition}
The base station is equipped with a UPA of dimension $M$ in the proposed system model. 
A digital beamforming design requires direct optimization of $M(N+1)$ complex weights, causing the curse of dimensionality.
To reduce the computational complexity, we restrict the beamforming vector at each time slot to lie in a subspace of rank $r$, which can be expressed as
\begin{equation}
\label{Kronecker}
\begin{aligned}
\boldsymbol{w}  = \sum_{k=1}^{r} \boldsymbol{u}_k \otimes \boldsymbol{v}_k,
\end{aligned}
\end{equation}
where the decomposition splits the two-dimensional directionality of the beamforming vector into independent one-dimensional factors of columns $\boldsymbol{u}_k$ and rows $\boldsymbol{v}_k$, respectively. Therefore, the number of learning parameters is reduced from $M(N+1)$ to $(M_x + M_y)(N+1)r$.
The representational capacity of the Kronecker parametrization is controlled by the rank parameter $r$. Increasing $r$ expands the set of feasible beamforming solutions when considering more complex scenarios, bringing the design closer to unconstrained digital beamforming, but at the cost of a larger action space and higher learning complexity, which makes the model training more difficult.

\subsubsection{QR Decomposition}
After performing the Kronecker decomposition, we apply QR decomposition to the beamforming vectors to enforce orthogonality among beams and thereby further suppress inter-user interference, which is expressed as
\begin{equation}
\label{QR}
\begin{aligned}
\mathbf{W}  = \mathbf{W}_Q \mathbf{W}_R, \;\; \mathbf{W} =\left[\boldsymbol{w}_{1,{t+1}},\dots,\boldsymbol{w}_{N,{t+1}},  \boldsymbol{w}_{0,{t+1}} \right],
\end{aligned}
\end{equation}
where $\mathbf{W}$ denotes the beamforming matrix, $\mathbf{W}_Q$ represents the orthonormalized beamforming matrix obtained via QR decomposition, and $\mathbf{W}_R$ is an upper triangular matrix, whose diagonal elements specify the transmit power allocated to each beamforming vector.

\subsection{C-HAPPO Algorithm Workflow}
Fig.~\ref{AlgorithmFramework} illustrates the C-HAPPO algorithm framework. Curriculum learning first categorizes tasks by adjusting grading parameters. In a curriculum task, the UAV and BS actor networks generate actions based on their observations in each time slot. The UAV outputs its trajectory for the next time slot, while the base station outputs beamforming factors via Kronecker decomposition, where these factors are then reassembled into beamforming vectors and orthonormalized. Finally, the UAV trajectory and beamforming vectors are applied to the environment to update the state. Once the cumulative reward exceeds a predefined threshold, curriculum learning increases the grading parameters to raise task difficulty, continuing until they reach the original values defined in the system model. During each interaction, experience tuples $\left\{\boldsymbol{s}_t,\boldsymbol{o}_t,\boldsymbol{a}_t,r(t),\boldsymbol{s}_{t+1} \right\}$ are stored in the experience replay buffer $\textbf{B}$. During each network update, batches of size $B$ are randomly sampled from this buffer to train both the critic and the actor networks. In the $n$-th actor network training, randomly select an agent to update the network sequentially, and then update the iterative equation $M^{n+1}\left(s_{}, \boldsymbol{a}_{}\right)$ for the next actor network update. The procedure of the C-HAPPO algorithm is outlined in Algorithm \ref{CHAPPOAlgorithmProcedure}.
	\begin{algorithm}[!t]
		\renewcommand{\algorithmicrequire}{\textbf{Input:}}
		\renewcommand{\algorithmicensure}{\textbf{Output:}}
		\caption{C-HAPPO Algorithm}
		\label{CHAPPOAlgorithmProcedure}
		\begin{algorithmic}[1]
			\STATE Set up the number of episodes $\mathcal{K}$, total time slot $T$, curriculum learning task number $L$, mini-batch size $B$.
			\STATE Initialize network parameters: actor networks $\vartheta_0$, critic network $\delta_0$, grading parameters $\mathfrak{R}_0$.
			\FOR{episode $k =  1 \to \mathcal{K}$}
			\FOR{time slot $t =  0 \to T$} 
			\STATE Output UAV actions $\boldsymbol{a}_{u_n,t}$ and BS action $\boldsymbol{a}_{bs,t}$ based on Eq.~(\ref{Kronecker}) and Eq.~(\ref{QR});
			\STATE Execute action $\boldsymbol{a}_t=(\boldsymbol{a}_{u_n,t},\boldsymbol{a}_{bs,t})$;
			\STATE Obtain the reward $r(t)$, the next state $\boldsymbol{s}_{t+1}$ and the next observation $\boldsymbol{o}_{t+1}$;
			\STATE Store the experience tuple $(\boldsymbol{s}_t,\boldsymbol{o}_{t}, \boldsymbol{a}_t,r(t),\boldsymbol{s}_{t+1})$ in the experience replay buffer $\textbf{B}$;
			\ENDFOR
			\STATE Calculate the cumulative reward ${R}_{}$;
			\IF {${R}_{} > {R}_{\text{min}}$}
			\STATE Update to the next curriculum learning task based on Eq.~(\ref{GradingParametersUpate});
			\ENDIF
			\STATE Sample a mini-batch $B$ from buffer $\textbf{B}$;
			\STATE Compute the advantage function $\hat{A}(s,\boldsymbol{a})$ based on the critic network;
			\STATE Set $M^{{1}}\left(s_{}, \boldsymbol{a}_{}\right) =\hat{A}(s,\boldsymbol{a})$;
			\FOR{Randomly select agent $n = 1 \to N+1$}
			\STATE Update the $n$-th actor network based on Eq. (\ref{ActorUpdate});
			\STATE Calculate $M^{{n+1}}\left(s_{}, \boldsymbol{a}_{}\right)$ based on Eq. (\ref{IterativeEquation});
			\ENDFOR
			\STATE Update the critic network based on Eq. (\ref{CriticUpdate});
			\ENDFOR
		\end{algorithmic}
	\end{algorithm}

\subsection{Computational Complexity Analysis}

We analyze the computational overhead of the proposed C-HAPPO algorithm in terms of training and inference time complexity. Let $H$ denote the number of neurons per hidden layer and $L_{\mathrm{nn}}$ the number of hidden layers in both the actor and critic networks.

\textit{Training Complexity:}
Each training episode consists of a data collection stage and a network update stage. During data collection, at each time slot, all $(N+1)$ agents perform forward passes through their actor networks, each incurring a cost of $\mathcal{O}(L_{\mathrm{nn}}H^2)$. The BS agent additionally performs Kronecker reconstruction at a cost of $\mathcal{O}(rM(N+1))$ and QR decomposition of the $M \times (N+1)$ beamforming matrix at a cost of $\mathcal{O}(M(N+1)^2)$. In the network update stage, the critic is trained over a mini-batch of $B$ samples at a cost of $\mathcal{O}(BTL_{\mathrm{nn}}H^2)$, and the $(N+1)$ actor networks are updated sequentially following the HAPPO mechanism, yielding $\mathcal{O}((N+1)BTL_{\mathrm{nn}}H^2)$. Aggregating both stages over $\mathcal{K}$ episodes, the overall training complexity is

\begin{equation}
\begin{aligned}
    \mathcal{O}_{\mathrm{train}} = \mathcal{O}\bigg(\mathcal{K} BT\Big[& \big((N\!+\!1)L_{\mathrm{nn}}H^2 + rM(N\!+\!1) \\
    &+ M(N\!+\!1)^2\big) + (N\!+\!2)L_{\mathrm{nn}}H^2 \Big] \bigg).
\end{aligned}
\end{equation}

\textit{Inference Complexity:}
During decentralized execution, each agent selects actions from local observations without accessing the critic network. At each time slot, all $(N+1)$ agents each perform a forward pass at $\mathcal{O}(L_{\mathrm{nn}}H^2)$, and the BS agent additionally carries out Kronecker reconstruction and QR decomposition. The per-slot inference complexity is therefore
\begin{equation}
    \mathcal{O}_{\mathrm{infer}} = \mathcal{O}\big((N\!+\!1)L_{\mathrm{nn}}H^2 + rM(N\!+\!1) + M(N\!+\!1)^2\big).
\end{equation}

\section{Simulation Results and Analysis}
In this section, we evaluate the performance of the proposed C-HAPPO algorithm through numerical simulations on the proposed system model. We first describe the simulation setup, then present the simulation results and analysis.

\subsection{Simulation Setup}

The simulation parameters of the proposed system model are summarized in Table \ref{SystemParameters}. The simulation is conducted in a three-dimensional Cartesian coordinate system, where the base station is located at $\boldsymbol{u}_{bs}=\left[0\,\text{m},0\,\text{m},0\,\text{m}\right]$, and the number of UAVs is set to $N=2$, where the locations of UAVs are $\boldsymbol{u}_{1_0}=\left[30\,\text{m}, 10\,\text{m}, 70\,\text{m}\right]$ and $\boldsymbol{u}_{2_0}=\left[30\,\text{m}, -10\,\text{m}, 70\,\text{m}\right]$, respectively. To guarantee the safety of the UAVs, the maximum speed of the UAVs is set to $\Delta u_{\text{max}}=5\,\text{m}/\text{s}$ and the minimum distance between UAVs is $ \Delta d_{\text{min}}=5\,\text{m}$. The sensed target is initially located at $\boldsymbol{\xi}_{0}=\left[60\,\text{m}, 0\,\text{m}, 60\,\text{m}\right]$, with initial velocity $\left[\dot{\xi}_{x,0},\dot{\xi}_{y,0},\dot{\xi}_{z,0}\right]  = \left[-1\,\text{m}/\text{s}, 1\,\text{m}/\text{s}, -1\,\text{m}/\text{s}\right]$.
For ground truth generation of the target, we utilize a constant acceleration model with acceleration $[0.02, -0.02, 0.02] \, \text{m}/\text{s}^2$ and a process noise intensity of $\sigma^2_{\zeta_t}=0.1 \, \text{m}^2/\text{s}^3$, a conservative value chosen to encompass the target's acceleration maneuvers within the tracker's noise covariance.
This produces a smooth but nontrivial trajectory that challenges the sensing estimator while remaining within typical kinematic limits for a slow-moving aerial target.
% The initial locations of the BS, UAVs, and the sensed target are shown in Fig.~\ref{InitalLocation}.
% \begin{figure}
% \centering
% \includegraphics[width=3.4in]{Figure/InitalLocation.pdf}
% \caption{The initial locations of the BS, UAVs, and the sensed target.}
% \label{InitalLocation}
% \end{figure}

In terms of the ISAC parameter settings, the base station is equipped with a uniform planar array with $M=8 \times 8$ antennas, and the bandwidth of the ISAC signals is set to $b=100\,\text{MHz}$. The transmit power of the base station $P$ is set to $5$ W, and the minimum communication SINR threshold is set to $5$ dB. The channel power gain at the reference distance is set to $\beta_{0}^{c}=-50$ dB for communication and $\beta_{0}^{s}=-50$ dB for sensing. 
The Rician factor $\Bbbk$ is set to 10. The CRB correlation constants are set to $\kappa_{d}=1$, $\kappa_{\theta}=10^{-6}$, $\kappa_{\varphi}=10^{-6}$, and $\kappa_{v}=10^{-6}$. For the distributed fusion among $N$ UAVs and one BS, we adopt Metropolis CI weights, where each weight is set to $\omega_n = \frac{1}{N+1}$.
The noise power $\sigma^2_0$, $\sigma^2_{u_{\text{bs}}}$, and $\sigma^2_{u_{n}}$ is set to $-80$ dBm, $-70$ dBm, and $-70$ dBm, respectively. The RCS of the sensed target is set to $0.9$. To verify the long-term stability of the system, the total time slot is set to $T=100$ s, with each time slot lasting 1 s.

\begin{table}
\centering
\caption{Simulation Parameters in the Proposed System Model \cite{dong2022sensing,zhu2024collaborative,meredith2015technical,xiao2005scheme}.}
\label{SystemParameters}
\resizebox{3.4in}{!}{
	\begin{tabular}{|c|c|}
		\hline Location of BS $\boldsymbol{u}_{bs}$ & $\left[0\,\text{m},0\,\text{m},0\,\text{m}\right]$\\
		\hline Number of UAVs $N$ & $2$\\
		\hline Initial location of UAV 1 $\boldsymbol{u}_{1_0}$ & $\left[30\,\text{m}, 10\,\text{m}, 70\,\text{m}\right]$\\
		\hline Initial location of UAV 2 $\boldsymbol{u}_{2_0}$ & $\left[30\,\text{m}, -10\,\text{m}, 70\,\text{m}\right]$\\
		\hline Maximum velocity of UAVs $ \Delta u_{\text{max}}$ & $5\,\text{m}/\text{s}$\\
		\hline Minimum distance between UAVs $ \Delta d_{\text{min}}$ & $5\,\text{m}$\\
		\hline Initial location of the sensed target $\boldsymbol{\xi}_0$ & $\left[60\,\text{m}, 0\,\text{m}, 60\,\text{m}\right]$\\
		\hline Initial velocity of the sensed target $\left[\dot{\xi}_{x,0},\dot{\xi}_{y,0},\dot{\xi}_{z,0}\right]$ & $\left[-1\,\text{m}/\text{s}, 1\,\text{m}/\text{s}, -1\,\text{m}/\text{s}\right]$\\
		\hline Acceleration of the sensed target & $\left[ 0.02\,\text{m}/\text{s}^2, -0.02\,\text{m}/\text{s}^2, 0.02\,\text{m}/\text{s}^2\right]$\\
        \hline Process noise intensity $\sigma^2_{\zeta_t}$ & $0.1\,\text{m}^2/\text{s}^3$\\
		\hline Number of BS transmit antennas $M$ & $8 \times 8$\\
		\hline Bandwidth of ISAC signals $b$  & $100\,\text{MHz}$\\
		\hline Distance between adjacent antennas of BS $d_x, d_y$ & $\lambda/2$\\
		\hline BS transmit power $P$ & 5 W \\
		\hline Minimum communication SINR threshold $\gamma_{\text{min}}$ & 5 dB\\
		\hline Communication channel power gain at reference distance $\beta_{0}^{c}$ & $-50$ dB\\
		\hline Sensing channel power gain at reference distance $\beta_{0}^{s}$ & $-50$ dB\\
        \hline  Rician factor $\Bbbk$ & 10\\
		\hline CRB correlation constants $\kappa_{d}$, $\kappa_{\theta}$, $\kappa_{\varphi}$, and $\kappa_{v}$ & $\left[1, 10 ^{-6}, 10 ^{-6}, 10 ^{-6}\right] $\\
        \hline CI fusion weight $\omega_n$ & $\frac{1}{N+1}$\\
		\hline Noise power $\sigma^2_0$ $\sigma^2_{u_{\text{bs}}}$, $\sigma^2_{u_{n}}$ & $-80$ dBm, $-70$ dBm, $-70$ dBm\\
		\hline RCS of the sensed target and of the UAVs $\alpha_0$, $\alpha_1$ & 0.9\\
		\hline Total time slots $T$ & 100 s\\
		\hline Duration of each time slot  & 1 s\\
		\hline 
	\end{tabular}}
\end{table}

The neural network and training parameters used in the proposed C-HAPPO algorithm are summarized in Table \ref{CHAPPOParameters}. In the simulation, five parallel threads are used, each with an episode horizon of $T$, and the mini-batch size is set to $5$. The proposed C-HAPPO algorithm comprises two-layer fully connected neural networks for both the actor and critic networks, where each hidden layer has 512 neurons. The learning rate of the networks is $0.0001$. 
The reward weighting coefficients are $\lambda^s=1$ and $\lambda^c=5$. The cumulative reward threshold is set to ${R}_{\text{min}}=-800$, and the total number of curriculum learning tasks is $L=2$. The reduced grading parameters are set to $\mathfrak{R}_{1} = \left\{1\,\text{dB}, 0.2, 2*10^{-7}, 2*10^{-7}, 2*10^{-7}\right\}$, and the Kronecker product decomposition rank is $r=2$.
\begin{table}[!t]
	\centering
	\caption{Simulation Parameters in the Proposed C-HAPPO Algorithm}
	\label{CHAPPOParameters}
	\resizebox{3.4in}{!}{
		\begin{tabular}{|c|c|}
			%			\hline \text { Parameter } & \text { Value } \\
            \hline Number of parallel threads & 5\\
            \hline Number of mini-batch size & 5\\
			\hline Number of hidden layers & 2\\
			\hline Number of neurons in each hidden layer & 512\\
			\hline Learning rate of network & 0.0001 \\
			\hline Reward discount factor $\gamma$ & 0.99\\
			\hline Reward weighting coefficients $\lambda^s,\lambda^c$ & $\left[1,5\right]$\\
			\hline Cumulative reward threshold ${R}_{\text{min}}$ & -800\\
			\hline Curriculum learning task number $L$ & 2\\ 
			\hline Reduced grading parameters $\mathfrak{R}_{1}$ & $\left\{1\,\text{dB}, 0.2, 2*10^{-7}, 2*10^{-7}, 2*10^{-7}\right\}$\\
			\hline Kronecker decomposition rank $r$ & 2\\
			\hline
	\end{tabular}}
\end{table}

To evaluate the performance of the proposed algorithm, we compare it with the following schemes as benchmarks:
\begin{itemize}
	\item \textbf{HAPPO}: Heterogeneous-agent proximal policy optimization algorithm (HAPPO) \cite{zhong2024heterogeneous} represents a state-of-the-art MADRL approach, which is designed for heterogeneous action space problems. It serves as a baseline for comparison with the proposed algorithm.
	\item \textbf{GA}: The genetic algorithm (GA) \cite{holland1992genetic}, a traditional optimization technique, is included as a benchmark for comparison of the proposed MADRL algorithm against optimization techniques.
	\item \textbf{RANDOM}: Each agent independently samples its action from a uniform distribution over the corresponding feasible action space.
	\item \textbf{NCL}: No curriculum learning (NCL) represents the algorithm having the curriculum learning removed from the proposed C-HAPPO, which is used to compare the effect of curriculum learning in the ablation experiment.
	\item \textbf{NKR}: No Kronecker decomposition (NKR) represents the algorithm having the Kronecker decomposition removed from the proposed C-HAPPO, which is used to compare the effect of Kronecker decomposition in the ablation experiment.
	\item \textbf{NQR}: No QR decomposition (NQR) represents the algorithm having the QR decomposition removed from the proposed C-HAPPO, which is used to compare the effect of QR decomposition in the ablation experiment.
\end{itemize}

\subsection{Simulation Results and Analysis}

\begin{figure}[!t]
\centering
\includegraphics[width=3in]{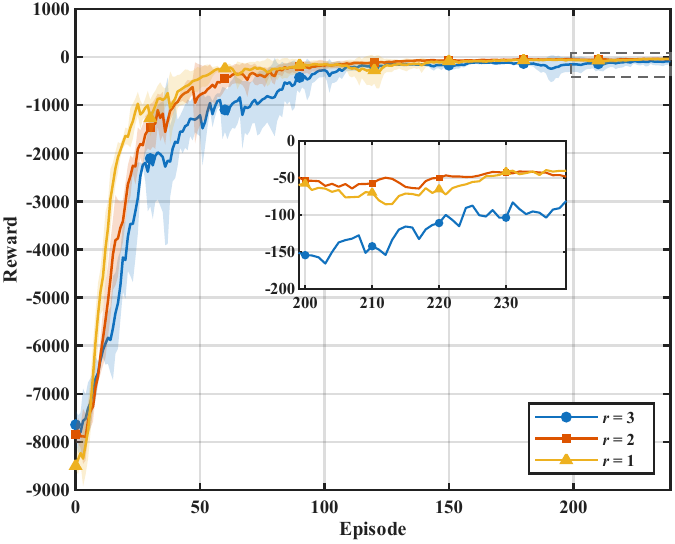}
\caption{Reward performance of the C-HAPPO algorithm with different Kronecker decomposition ranks $r$.}
\label{KRCurve}
\end{figure}

Firstly, we compare the reward performance under different Kronecker decomposition ranks $r$ in the proposed C-HAPPO algorithm, as shown in Fig.~\ref{KRCurve}. It can be seen that after convergence, the average reward of $r = 2$ is slightly higher than that of $r = 1$ and significantly outperforms $r = 3$. Although $r=1$ converges quickly, $r=2$ ultimately achieves a superior stable reward. This indicates that while a larger $r$ (e.g., $r=3$) increases the learning parameters and makes training difficult, $r = 2$ provides a better balance between representational capability and training efficiency compared to $r=1$. Hence, a Kronecker decomposition rank of $r = 2$ is selected in the proposed C-HAPPO algorithm for the subsequent simulations.

\begin{figure}[!t]
\centering
\includegraphics[width=3in]{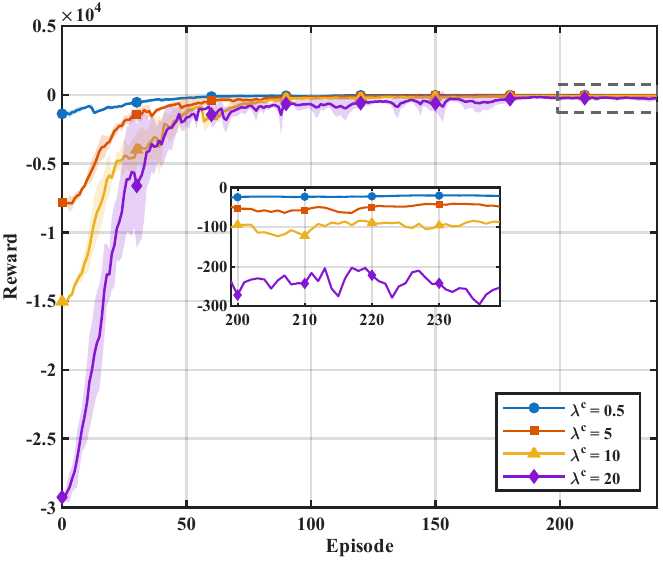}
\caption{Reward performance of the C-HAPPO algorithm under different communication penalty coefficients $\lambda^c$.}
\label{RewardCurve_lambda}
\end{figure}

\begin{figure}[!t]
\centering
\includegraphics[width=3in]{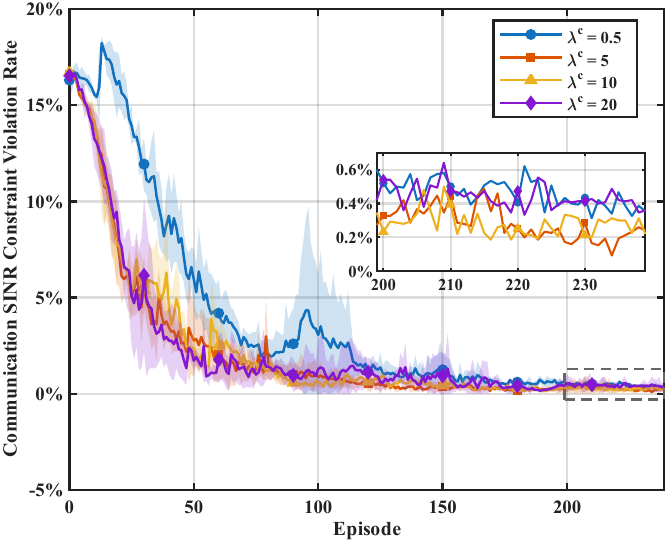}
\caption{Communication SINR constraint violation rate under different communication penalty coefficients $\lambda^c$.}
\label{ViolationCurve_lambda}
\end{figure}

To determine an appropriate communication penalty coefficient $\lambda^c$, we evaluate the proposed method under various coefficient values. As illustrated in Fig.~\ref{RewardCurve_lambda}, setting a small coefficient ($\lambda^c=0.5$) yields the highest reward, as the agent prioritizes maximizing the primary objective over penalizing communication violations. However, Fig.~\ref{ViolationCurve_lambda} indicates that this configuration results in a relatively high residual communication SINR constraint violation rate. Conversely, increasing $\lambda^c$ to 10 or 20 makes the communication penalty overly dominant. This suppresses exploration flexibility and substantially degrades the final reward, while failing to further reduce the residual violation rate compared to $\lambda^c=5$. Therefore, we adopt $\lambda^c=5$ for the proposed C-HAPPO algorithm, as it achieves an optimal trade-off between system reward maximization and communication constraint satisfaction.

\begin{figure}[!t]
\centering
\includegraphics[width=3.1in]{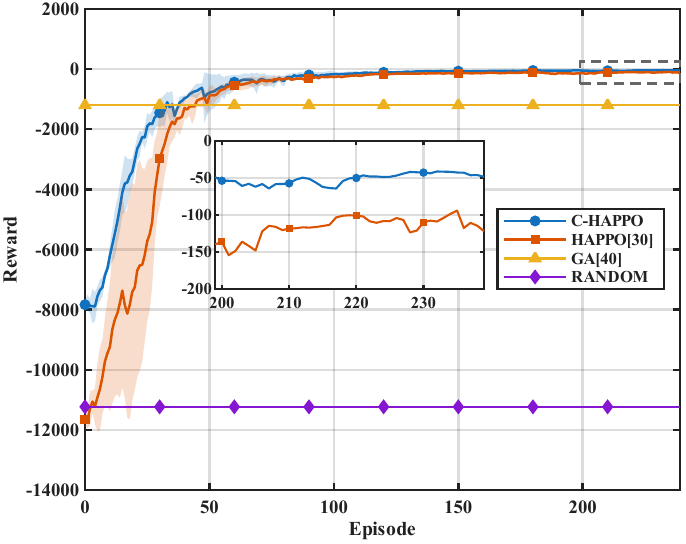}
\caption{Reward convergence of the C-HAPPO algorithm compared with benchmark methods.}
\label{RewardCurve}
\end{figure}

Fig.~\ref{RewardCurve} reports the reward convergence of the proposed C-HAPPO algorithm with the benchmark schemes. It can be seen that the proposed C-HAPPO algorithm achieves faster convergence than the HAPPO algorithm, where the C-HAPPO algorithm converges after 50 episodes, while the HAPPO algorithm converges after 90 episodes. In addition, the proposed C-HAPPO algorithm achieves the highest average reward of about $-50$, which is significantly higher than those achieved by GA, HAPPO, and RANDOM.

\begin{figure}[!t]
\centering
\includegraphics[width=3in]{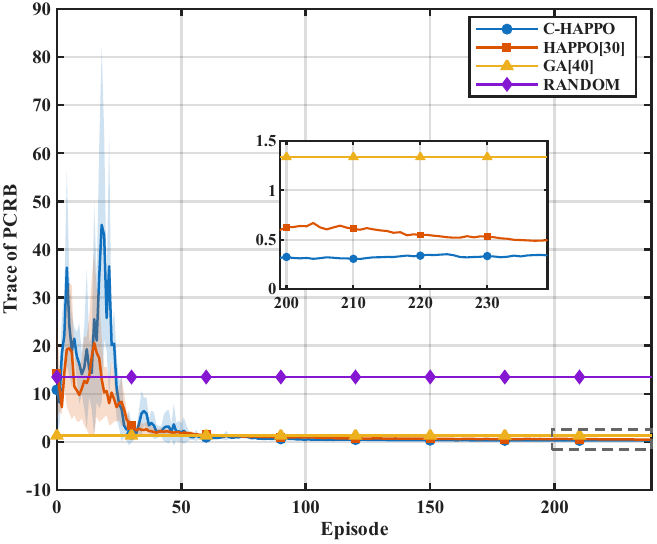}
\caption{Average sensing performance of the C-HAPPO algorithm compared with benchmark methods.}
\label{PCRBCurve}
\end{figure}

\begin{figure*}[!t]
\centering
\subfigure[]{
    \includegraphics[width=0.3\textwidth]{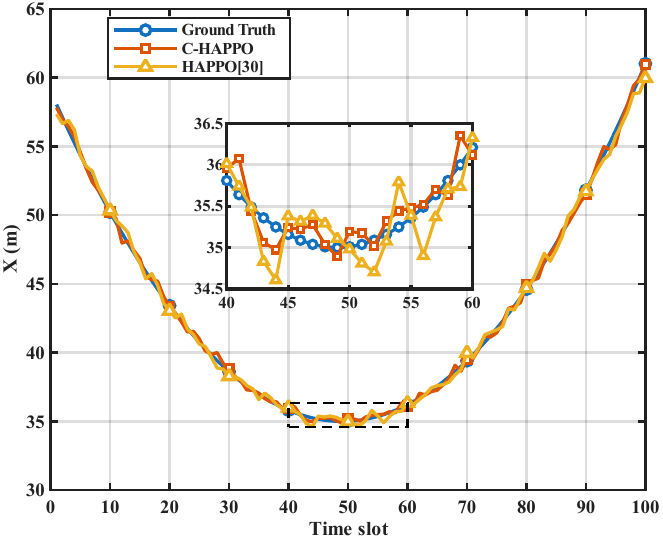}
}
\hfill  
\subfigure[]{
    \includegraphics[width=0.3\textwidth]{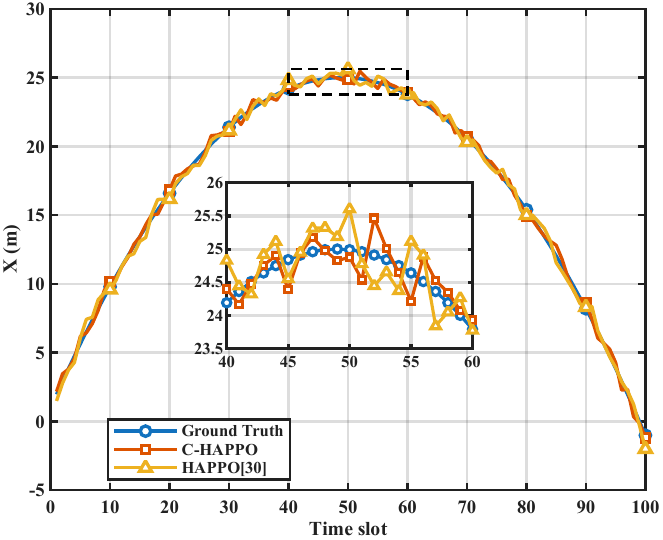}
}
\hfill 
\subfigure[]{
    \includegraphics[width=0.3\textwidth]{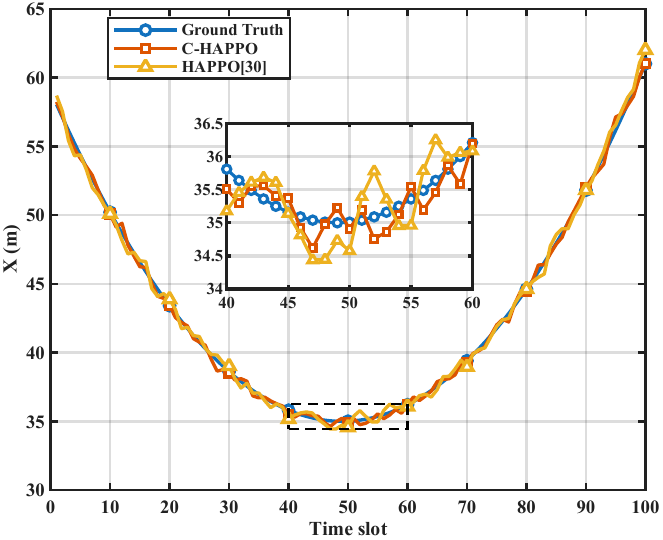}
}
\caption{Comparison of predicted and actual locations of the sensed target with different methods: $x$-axis (a), $y$-axis (b), and $z$-axis (c).}
\label{2DTargetFlightCurve}
\end{figure*}

\begin{figure}[!htbp]
\centering
\includegraphics[width=3in]{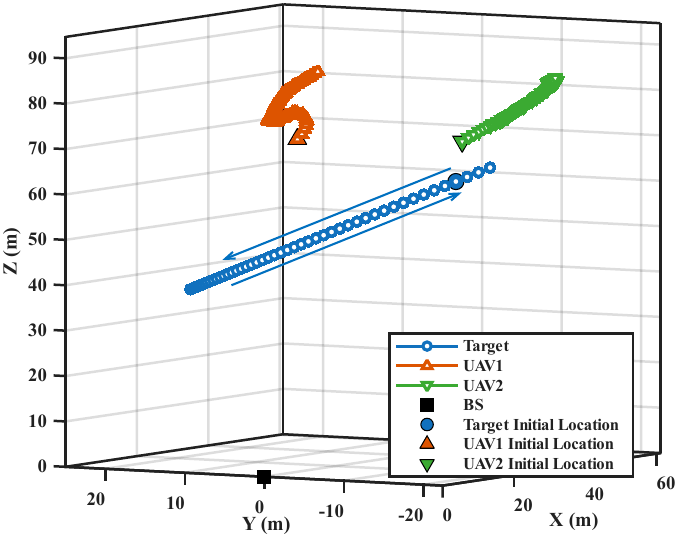}
\caption{Flight trajectories of UAVs and the sensed target of the C-HAPPO algorithm for the proposed system model.}
\label{3DFlightCurve}
\end{figure}

\begin{figure}[!t]
\centering
\includegraphics[width=3in]{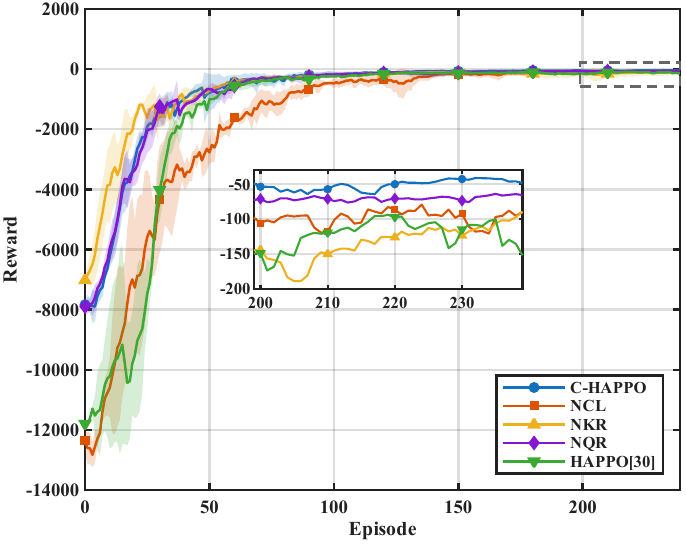}
\caption{Ablation experiment performance analysis for reward.}
\label{AblationRewardCurve}
\end{figure}

\begin{figure}[!t]
\centering
\includegraphics[width=3in]{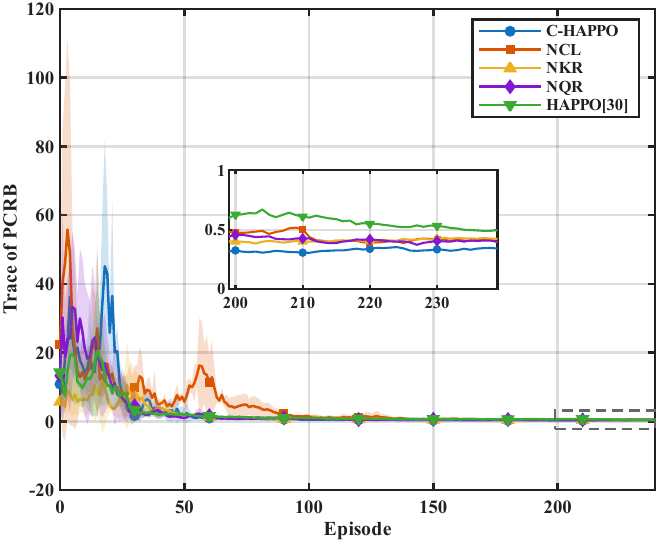}
\caption{Ablation experiment performance analysis for PCRB.}
\label{AblationPCRBCurve}
\end{figure}

% \begin{figure}[!htbp]
% \centering
% \subfigure[]{
% \includegraphics[width=3in]{Figure/TargetLocationPredictXCurve.pdf}
% }
% \subfigure[]{
% \includegraphics[width=3in]{Figure/TargetLocationPredictYCurve.pdf}
% }
% \subfigure[]{
% \includegraphics[width=3in]{Figure/TargetLocationPredictZCurve.pdf}
% }
% \caption{Comparison of predicted and actual locations of the sensed target with different methods: $x$-axis (a), $y$-axis (b), and $z$-axis (c).}
% \label{2DTargetFlightCurve}
% \end{figure}

\begin{figure}[!htbp]
\centering
\includegraphics[width=3in]{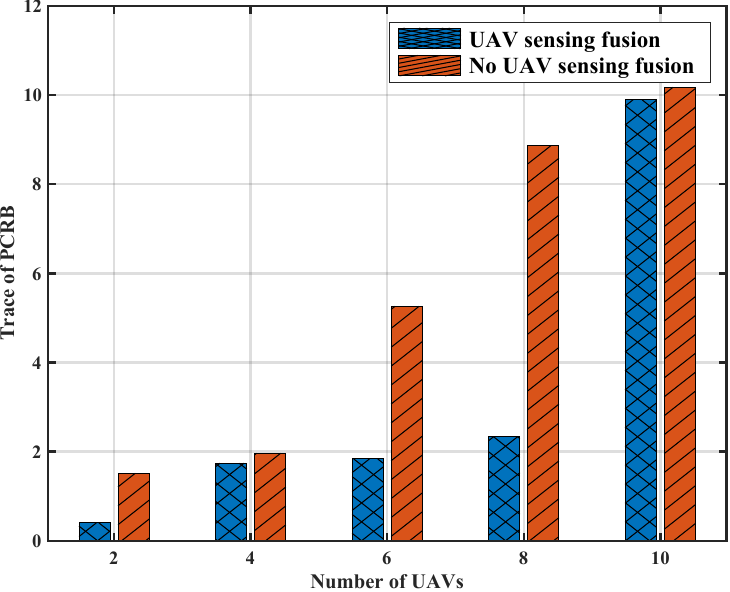}
\caption{Sensing performance under different number of UAVs.}
\label{SensingFuseCurve}
\end{figure}

Similarly, we compare the average PCRB performance of the proposed C-HAPPO algorithm with the benchmark methods in Fig.~\ref{PCRBCurve}. After 60 episodes, the trace of PCRB of the C-HAPPO algorithm remains consistently lower than that of the other benchmark algorithms. After all algorithms have converged, the trace of PCRB of the C-HAPPO algorithm remains around 0.35, while that of the GA algorithm is 1.34, that of the HAPPO algorithm remains above 0.53, and that of the RANDOM method is 13.5. Therefore, compared with the reference algorithms, the proposed C-HAPPO algorithm achieves over 30\% improvement in sensing performance.

Fig.~\ref{2DTargetFlightCurve} demonstrates the trajectory of the sensed target with the tracking results of two methods, C-HAPPO and HAPPO, over 100 time slots, based on the $x$, $y$, and $z$ coordinate components, respectively. It can be observed that the predicted trajectory of the C-HAPPO algorithm closely matches the Ground Truth throughout the entire flight process. In time slots 40–60, the C-HAPPO almost coincides with the true trajectory. In contrast, while the HAPPO algorithm generally follows the target's movement, it exhibits noticeable fluctuations and larger tracking errors around the Ground Truth. Consequently, the C-HAPPO algorithm demonstrates significantly higher tracking accuracy and stability compared to HAPPO.

Fig.~\ref{3DFlightCurve} illustrates the flight trajectories of the UAVs and the sensed target in the proposed system model. The dynamic interaction is divided into two phases based on the target's motion. In the phase where the target moves downwards, UAV 1 performs local maneuvers to enhance sensing performance, while UAV 2 moves towards the negative Y-direction. The movement of UAV 2 widens the angular diversity relative to the BS, effectively reducing inter-beam coupling. Subsequently, in the phase where the target moves upwards, both UAV 1 and UAV 2 synchronize their flight directions with the target to maintain stable tracking and communication.

Fig.~\ref{AblationRewardCurve} and Fig.~\ref{AblationPCRBCurve} illustrate the convergence processes of the five algorithms within 240 training rounds to evaluate the contributions of the three proposed improvements, including curriculum learning (CL), Kronecker decomposition (KR), and QR decomposition (QR), to the original HAPPO.
In Fig.~\ref{AblationRewardCurve}, it is evident that the proposed C-HAPPO algorithm exhibits a superior convergence rate compared to the standard HAPPO benchmark and the ablation variants (NCL, NKR, and NQR). While all schemes eventually trend towards stability, C-HAPPO achieves a high-value policy rapidly, stabilizing approximately after 50 episodes. In contrast, the baseline algorithms suffer from slower learning curves and noticeable fluctuations during the early training stages. The final episodes (Episodes 200–240) further highlight that C-HAPPO maintains the highest reward. This performance gap indicates that the proposed mechanisms effectively mitigate local optima, enabling the agents to learn a more robust cooperative policy for the ISAC task.
In terms of the sensing performance, Fig.~\ref{AblationPCRBCurve} evaluates the sensing performance in terms of the Trace of PCRB. The results demonstrate that the proposed C-HAPPO algorithm consistently achieves the lowest PCRB among all compared methods, stabilizing at a value below 0.35 in the final episodes. Conversely, the HAPPO benchmark exhibits the highest PCRB (0.53), reflecting limited sensing capability in dynamic environments. The ablation variants (NCL, NKR, and NQR) lie between these two extremes, with their performance degradation confirming that each removed module is critical for minimizing the estimation error. This suggests that the trajectory planning and resource allocation strategies optimized by C-HAPPO successfully maximize the sensing gain.

Fig.~\ref{SensingFuseCurve} characterizes the sensing performance trends of two strategies, UAV sensing fusion and no UAV sensing fusion, for different numbers of UAVs. It is evident that the UAV sensing fusion scheme consistently maintains a lower PCRB compared to the non-fusion baseline regardless of the number of UAVs, validating the robustness of the proposed fusion mechanism. The performance advantage is most significant with 8 UAVs, where the non-fusion approach suffers from a sharp degradation in sensing accuracy (rising to 9.0). In contrast, the UAV sensing fusion scheme effectively suppresses the estimation error, maintaining PCRB below 2.5 with 8 UAVs, by leveraging complementary information shared among agents. Although the global PCRB naturally increases with the network size due to the escalating complexity of the larger-scale network environment and potential inter-UAV interference, the proposed method demonstrates superior resilience against these scaling challenges compared to the baseline.

% \begin{figure}
% \centering
% \includegraphics[width=\columnwidth]{Figure/Algorithm.eps}
% \caption{Operation process of the C-HAPPO algorithm for the proposed system model.}
% \label{NoUAVSensingCurve}
% \end{figure}

% \begin{figure}
% \centering
% \includegraphics[width=\columnwidth]{Figure/Algorithm.eps}
% \caption{Operation process of the C-HAPPO algorithm for the proposed system model.}
% \label{DifferentUAVNumberCurve}
% \end{figure}

\section{Conclusions}

This paper introduces a UAV-BS ISAC system model, where the UAVs are tasked with sensing a moving target while maintaining communication with the base station. To minimize the average PCRB of the sensed target, we formulate a joint beamforming and trajectory optimization problem.
We propose a C-HAPPO algorithm to address the challenges of joint beamforming and trajectory optimization. Simulation results demonstrate that the proposed C-HAPPO algorithm outperforms benchmark methods in terms of reward and average PCRB. 

For future research, there are several promising extensions. First, integrating the C-HAPPO algorithm with Random Finite Set (RFS) theory will be investigated to address the coupled challenges of resource allocation and data association in multi-target tracking. In addition, the proposed BS-assisted cooperative ISAC framework can be extended to communication-limited scenarios by utilizing compressed observations or intermittent reporting. Developing fully distributed ISAC architectures with on-board interference cancellation and local sensing recovery capabilities also represents an important direction for future research.

\FloatBarrier            % 刷新所有浮动体
\footnotesize
\bibliography{reference}

\begin{thebibliography}{10}
\providecommand{\url}[1]{#1}
\csname url@samestyle\endcsname
\providecommand{\newblock}{\relax}
\providecommand{\bibinfo}[2]{#2}
\providecommand{\BIBentrySTDinterwordspacing}{\spaceskip=0pt\relax}
\providecommand{\BIBentryALTinterwordstretchfactor}{4}
\providecommand{\BIBentryALTinterwordspacing}{\spaceskip=\fontdimen2\font plus
\BIBentryALTinterwordstretchfactor\fontdimen3\font minus
  \fontdimen4\font\relax}
\providecommand{\BIBforeignlanguage}[2]{{%
\expandafter\ifx\csname l@#1\endcsname\relax
\typeout{** WARNING: IEEEtran.bst: No hyphenation pattern has been}%
\typeout{** loaded for the language `#1'. Using the pattern for}%
\typeout{** the default language instead.}%
\else
\language=\csname l@#1\endcsname
\fi
#2}}
\providecommand{\BIBdecl}{\relax}
\BIBdecl

\bibitem{wang2023road}
C.-X. Wang, X.~You, X.~Gao, X.~Zhu, Z.~Li, C.~Zhang, H.~Wang, Y.~Huang,
  Y.~Chen, H.~Haas \emph{et~al.}, ``On the road to {6G}: Visions, requirements,
  key technologies, and testbeds,'' \emph{IEEE Communications Surveys \&
  Tutorials}, vol.~25, no.~2, pp. 905--974, 2023.

\bibitem{na2024operator}
M.~Na, J.~Lee, G.~Choi, T.~Yu, J.~Choi, J.~Lee, and S.~Bahk, ``Operator's
  perspective on {6G}: {6G} services, vision, and spectrum,'' \emph{IEEE
  Communications Magazine}, vol.~62, no.~8, pp. 178--184, 2024.

\bibitem{luo2025isac}
X.~Luo, Q.~Lin, R.~Zhang, H.-H. Chen, X.~Wang, and M.~Huang, ``{ISAC}--a survey
  on its layered architecture, technologies, standardizations, prototypes and
  testbeds,'' \emph{IEEE Communications Surveys \& Tutorials}, 2025.

\bibitem{luo2025bedrock}
C.~Luo, L.~Xiang, J.~Hu, and K.~Yang, ``Bedrock models in communication and
  sensing: Advancing generalization, transferability, and performance,''
  \emph{arXiv preprint arXiv:2503.08220}, 2025.

\bibitem{peng11165352}
Y.~Peng, L.~Xiang, K.~Yang, F.~Jiang, K.~Wang, and D.~O. Wu, ``Simac: A
  semantic-driven integrated multimodal sensing and communication framework,''
  \emph{IEEE Journal on Selected Areas in Communications}, pp. 1--1, 2025.

\bibitem{zhou11357472}
S.~Zhou, L.~Xiang, Y.~Wang, K.~Yang, K.~K. Wong, and C.-B. Chae, ``Extended
  target adaptive beamforming for isac: A perspective of predictive error
  ellipse,'' \emph{IEEE Transactions on Wireless Communications}, vol.~25, pp.
  10\,604--10\,617, 2026.

\bibitem{kaushik2024toward}
A.~Kaushik, R.~Singh, S.~Dayarathna, R.~Senanayake, M.~Di~Renzo, M.~Dajer,
  H.~Ji, Y.~Kim, V.~Sciancalepore, A.~Zappone \emph{et~al.}, ``Toward
  integrated sensing and communications for {6G}: Key enabling technologies,
  standardization, and challenges,'' \emph{IEEE Communications Standards
  Magazine}, vol.~8, no.~2, pp. 52--59, 2024.

\bibitem{wang2025toward}
Y.~Wang, G.~Sun, Z.~Sun, J.~Wang, J.~Li, C.~Zhao, J.~Wu, S.~Liang, M.~Yin,
  P.~Wang \emph{et~al.}, ``Toward realization of low-altitude economy networks:
  Core architecture, integrated technologies, and future directions,''
  \emph{arXiv preprint arXiv:2504.21583}, 2025.

\bibitem{javed2024state}
S.~Javed, A.~Hassan, R.~Ahmad, W.~Ahmed, R.~Ahmed, A.~Saadat, and M.~Guizani,
  ``State-of-the-art and future research challenges in {UAV} swarms,''
  \emph{IEEE Internet of Things Journal}, vol.~11, no.~11, pp.
  19\,023--19\,045, 2024.

\bibitem{jiang2025integrated}
Y.~Jiang, X.~Li, G.~Zhu, H.~Li, J.~Deng, K.~Han, C.~Shen, Q.~Shi, and R.~Zhang,
  ``Integrated sensing and communication for low altitude economy:
  Opportunities and challenges,'' \emph{IEEE Communications Magazine}, 2025.

\bibitem{song2025overview}
Y.~Song, Y.~Zeng, Y.~Yang, Z.~Ren, G.~Cheng, X.~Xu, J.~Xu, S.~Jin, and
  R.~Zhang, ``An overview of cellular {ISAC} for low-altitude {UAV}: New
  opportunities and challenges,'' \emph{IEEE Communications Magazine}, 2025.

\bibitem{lyu2022joint}
Z.~Lyu, G.~Zhu, and J.~Xu, ``Joint maneuver and beamforming design for
  {UAV}-enabled integrated sensing and communication,'' \emph{IEEE Transactions
  on Wireless Communications}, vol.~22, no.~4, pp. 2424--2440, 2022.

\bibitem{deng2024joint}
D.~Deng, W.~Zhou, X.~Li, D.~B. Da~Costa, D.~W.~K. Ng, and A.~Nallanathan,
  ``Joint beamforming and {UAV} trajectory optimization for covert
  communications in {ISAC} networks,'' \emph{IEEE Transactions on Wireless
  Communications}, 2024.

\bibitem{meng2022throughput}
K.~Meng, Q.~Wu, S.~Ma, W.~Chen, K.~Wang, and J.~Li, ``Throughput maximization
  for {UAV}-enabled integrated periodic sensing and communication,'' \emph{IEEE
  Transactions on Wireless Communications}, vol.~22, no.~1, pp. 671--687, 2022.

\bibitem{jing2024isac}
X.~Jing, F.~Liu, C.~Masouros, and Y.~Zeng, ``{ISAC} from the sky: {UAV}
  trajectory design for joint communication and target localization,''
  \emph{IEEE Transactions on Wireless Communications}, vol.~23, no.~10, pp.
  12\,857--12\,872, 2024.

\bibitem{zhou2025beamforming}
S.~Zhou, L.~Xiang, K.~Yang, K.~K. Wong, D.~O. Wu, and C.-B. Chae,
  ``Beamforming-based achievable rate maximization in {ISAC} system for
  multi-{UAV} networking,'' \emph{arXiv preprint arXiv:2507.21895}, 2025.

\bibitem{diaz2023sensing}
C.~Diaz-Vilor, M.~A. Almasi, A.~M. Abdelhady, A.~Celik, A.~M. Eltawil, and
  H.~Jafarkhani, ``Sensing and communication in {UAV} cellular networks: Design
  and optimization,'' \emph{IEEE Transactions on Wireless Communications},
  vol.~23, no.~6, pp. 5456--5472, 2023.

\bibitem{liu2024uav}
Z.~Liu, X.~Liu, Y.~Liu, V.~C. Leung, and T.~S. Durrani, ``{UAV} assisted
  integrated sensing and communications for internet of things: {3D} trajectory
  optimization and resource allocation,'' \emph{IEEE Transactions on Wireless
  Communications}, vol.~23, no.~8, pp. 8654--8667, 2024.

\bibitem{khalili2024efficient}
A.~Khalili, A.~Rezaei, D.~Xu, F.~Dressler, and R.~Schober, ``Efficient {UAV}
  hovering, resource allocation, and trajectory design for {ISAC} with limited
  backhaul capacity,'' \emph{IEEE Transactions on Wireless Communications},
  2024.

\bibitem{wang2024isac}
Y.~Wang, K.~Zu, L.~Xiang, Q.~Zhang, Z.~Feng, J.~Hu, and K.~Yang, ``{ISAC}
  enabled cooperative detection for cellular-connected {UAV} network,''
  \emph{IEEE Transactions on Wireless Communications}, 2024.

\bibitem{garcia2013markov}
F.~Garcia and E.~Rachelson, ``Markov decision processes,'' \emph{Markov
  Decision Processes in Artificial Intelligence}, pp. 1--38, 2013.

\bibitem{li2023radio}
Y.~Li and A.~H. Aghvami, ``Radio resource management for cellular-connected
  {UAV}: A learning approach,'' \emph{IEEE Transactions on Communications},
  vol.~71, no.~5, pp. 2784--2800, 2023.

\bibitem{li2022path}
Y.~Li, A.~H. Aghvami, and D.~Dong, ``Path planning for cellular-connected
  {UAV}: A {DRL} solution with quantum-inspired experience replay,'' \emph{IEEE
  Transactions on Wireless Communications}, vol.~21, no.~10, pp. 7897--7912,
  2022.

\bibitem{li2025energy}
Y.~Li, A.~Madhukumar, T.~Z.~H. Ernest, G.~Zheng, W.~Saad, and A.~H. Aghvami,
  ``Energy-efficient {UAV}-driven multi-access edge computing: a distributed
  many-agent perspective,'' \emph{IEEE Transactions on Communications}, 2025.

\bibitem{gao2024marl}
Q.~Gao, R.~Zhong, H.~Shin, and Y.~Liu, ``{MARL} based {UAV}s’ trajectory and
  beamforming optimization for {ISAC} system,'' \emph{IEEE Internet of Things
  Journal}, 2024.

\bibitem{xie2024distributed}
Z.~Xie, Z.~Wang, Z.~Zhang, J.~Wang, Z.~Jiang, and Z.~Han, ``Distributed {UAV}
  swarm for device-free integrated sensing and communication relying on
  multi-agent reinforcement learning,'' \emph{IEEE Transactions on Vehicular
  Technology}, 2024.

\bibitem{cheng2024joint}
S.~Cheng, X.~Lin, X.~Li, and J.~Wang, ``Joint {UAV} trajectory and radcom task
  schedule for {IVN}s: A game-embedding multi-agent deep reinforcement learning
  approach,'' \emph{IEEE Transactions on Wireless Communications}, 2024.

\bibitem{qin2023deep}
Y.~Qin, Z.~Zhang, X.~Li, W.~Huangfu, and H.~Zhang, ``Deep reinforcement
  learning based resource allocation and trajectory planning in integrated
  sensing and communications {UAV} network,'' \emph{IEEE Transactions on
  Wireless Communications}, vol.~22, no.~11, pp. 8158--8169, 2023.

\bibitem{ye2025aoi}
Y.~Ye, Y.~Tian, C.~H. Liu, L.~Dong, G.~Qi, and D.~Wu, ``{AoI}-aware air-ground
  mobile crowdsensing by multi-agent curriculum learning with collaborative
  observation augmentation,'' \emph{IEEE Transactions on Mobile Computing},
  no.~01, pp. 1--13, 2025.

\bibitem{zhong2024heterogeneous}
Y.~Zhong, J.~G. Kuba, X.~Feng, S.~Hu, J.~Ji, and Y.~Yang, ``Heterogeneous-agent
  reinforcement learning,'' \emph{Journal of Machine Learning Research},
  vol.~25, no.~32, pp. 1--67, 2024.

\bibitem{li2017joint}
B.~Li and A.~P. Petropulu, ``Joint transmit designs for coexistence of {MIMO}
  wireless communications and sparse sensing radars in clutter,'' \emph{IEEE
  Transactions on Aerospace and Electronic Systems}, vol.~53, no.~6, pp.
  2846--2864, 2017.

\bibitem{li2018optimal}
X.~Li, C.~Tepedelenlio{\u{g}}lu, and H.~{\c{S}}enol, ``Optimal training for
  residual self-interference for full-duplex one-way relays,'' \emph{IEEE
  Transactions on Communications}, vol.~66, no.~12, pp. 5976--5989, 2018.

\bibitem{dong2022sensing}
F.~Dong, F.~Liu, Y.~Cui, W.~Wang, K.~Han, and Z.~Wang, ``Sensing as a service
  in {6G} perceptive networks: A unified framework for {ISAC} resource
  allocation,'' \emph{IEEE Transactions on Wireless Communications}, vol.~22,
  no.~5, pp. 3522--3536, 2022.

\bibitem{liu2020radar}
F.~Liu, W.~Yuan, C.~Masouros, and J.~Yuan, ``Radar-assisted predictive
  beamforming for vehicular links: Communication served by sensing,''
  \emph{IEEE Transactions on Wireless Communications}, vol.~19, no.~11, pp.
  7704--7719, 2020.

\bibitem{blair2021industry}
W.~Blair, ``Industry tip: Picking the minimum process noise variance for your
  {NCV} track filter,'' \emph{IEEE Aerospace and Electronic Systems Magazine},
  vol.~36, no.~2, pp. 72--74, 2021.

\bibitem{schulman2015high}
J.~Schulman, P.~Moritz, S.~Levine, M.~Jordan, and P.~Abbeel, ``High-dimensional
  continuous control using generalized advantage estimation,'' \emph{arXiv
  preprint arXiv:1506.02438}, 2015.

\bibitem{zhu2024collaborative}
Y.~Zhu, M.~Chen, S.~Wang, Y.~Hu, Y.~Liu, and C.~Yin, ``Collaborative
  reinforcement learning based unmanned aerial vehicle ({UAV}) trajectory
  design for {3D} {UAV} tracking,'' \emph{IEEE Transactions on Mobile
  Computing}, vol.~23, no.~12, pp. 10\,787--10\,802, 2024.

\bibitem{meredith2015technical}
J.~Meredith, ``Technical specification group radio access network: Study on
  enhanced {LTE} support for aerial vehicles,'' 2015.

\bibitem{xiao2005scheme}
L.~Xiao, S.~Boyd, and S.~Lall, ``A scheme for robust distributed sensor fusion
  based on average consensus,'' in \emph{IPSN 2005. Fourth International
  Symposium on Information Processing in Sensor Networks, 2005.}\hskip 1em plus
  0.5em minus 0.4em\relax IEEE, 2005, pp. 63--70.

\bibitem{holland1992genetic}
J.~H. Holland, ``Genetic algorithms,'' \emph{Scientific american}, vol. 267,
  no.~1, pp. 66--73, 1992.

\end{thebibliography}

\end{document}